\documentclass[aps,prd,showkeys,superscriptaddress,twocolumn,nofootinbib,floatfix]{revtex4-1}

\pdfoutput=1

\usepackage{amsmath}
\usepackage{amssymb}
\usepackage{amsthm}
\usepackage{mathrsfs}
\usepackage{graphicx}
\usepackage{epstopdf}
\usepackage{fancyhdr}
\usepackage{array}
\usepackage[all]{xy}
\usepackage{eufrak}
\usepackage{euscript}
\usepackage{enumerate}
\usepackage{slashed}
\usepackage{hyperref}
\usepackage{subfigure} 
\usepackage{epstopdf} 

\hypersetup{pdftex,colorlinks=true,linkcolor=blue,citecolor=blue,menucolor=black,urlcolor=blue,filecolor=blue}

\begin{document}

\title{Critical point in the phase diagram of primordial quark-gluon matter from black hole physics}

\author{Renato Critelli}
\email{renato.critelli@usp.br}
\affiliation{Instituto de F\'{i}sica, Universidade de S\~{a}o Paulo, Rua do Mat\~{a}o, 1371, Butant\~{a}, CEP 05508-090, S\~{a}o Paulo, S\~{a}o Paulo, Brazil}

\author{Jorge Noronha}
\email{noronha@if.usp.br}
\affiliation{Instituto de F\'{i}sica, Universidade de S\~{a}o Paulo, Rua do Mat\~{a}o, 1371, Butant\~{a}, CEP 05508-090, S\~{a}o Paulo, S\~{a}o Paulo, Brazil}

\author{Jacquelyn Noronha-Hostler}
\email{jakinoronhahostler@gmail.com}
\affiliation{Department of Physics and Astronomy, Rutgers University, 136 Frelinghuysen Rd, Piscataway, NJ 08854, USA}
\affiliation{Department of Physics, University of Houston, Houston TX 77204, USA}

\author{Israel Portillo}
\email{iportillovazquez@gmail.com}
\affiliation{Department of Physics, University of Houston, Houston TX 77204, USA}

\author{Claudia Ratti}
\email{cratti@uh.edu}
\affiliation{Department of Physics, University of Houston, Houston TX 77204, USA}

\author{Romulo Rougemont}
\email{rrougemont@iip.ufrn.br}
\affiliation{International Institute of Physics, Federal University of Rio Grande do Norte,
Campus Universit\'{a}rio - Lagoa Nova, CEP 59078-970, Natal, Rio Grande do Norte, Brazil}

\begin{abstract}
Strongly interacting matter undergoes a crossover phase transition at high temperatures $T\sim 10^{12}$ K and zero net-baryon density. A fundamental question in the theory of strong interactions, Quantum Chromodynamics (QCD), is whether a hot and dense system of quarks and gluons displays critical phenomena when doped with more quarks than antiquarks, where net-baryon number fluctuations diverge. Recent lattice QCD work indicates that such a critical point can only occur in the baryon dense regime of the theory, which defies a description from first principles calculations. Here we use the holographic gauge/gravity correspondence to map the fluctuations of baryon charge in the dense quark-gluon liquid onto a numerically tractable gravitational problem involving the charge fluctuations of holographic black holes. This approach quantitatively reproduces ab initio results for the lowest order moments of the baryon fluctuations and makes predictions for the higher order baryon susceptibilities and also for the location of the critical point, which is found to be within the reach of heavy ion collision experiments.
\end{abstract}



\maketitle

\section{Introduction}
\label{sec:intro}

The rapid crossover transition found in lattice QCD calculations \cite{Aoki:2006we} characterizes the change in the degrees of freedom of the theory from hadrons to a novel deconfined state composed of quarks and gluons. The extreme conditions needed for this phenomenon took place in our Universe $\sim 20$ microseconds after the Big Bang \cite{Weinberg:2008zzc} and have been constantly reproduced over the last decade in ultrarelativistic heavy ion collisions at the Relativistic Heavy Ion Collider (RHIC) and the Large Hadron Collider (LHC). These experiments have provided overwhelming evidence that at high temperatures quarks and gluons form a new type of strongly interacting liquid called the quark-gluon plasma (QGP) \cite{Shuryak:1980tp}. Since its discovery in the early 2000's (see, e.g., Ref.\ \cite{Gyulassy:2004zy} for a review), it has become clear that the femtoscopic version of the primordial liquid recreated in these experiments possesses novel many-body properties, including nearly inviscid flow behavior characterized by a surprisingly small value \cite{Heinz:2013th} of its shear viscosity to entropy density ratio ($\eta/s$), which makes the QGP the smallest (and the hottest) most perfect fluid ever observed.

Despite the steady progress over the years in the determination of the QGP's equilibrium properties through lattice simulations, most regions of the QCD phase diagram remain vastly unexplored. In fact, ab initio calculations in the baryon dense regime of QCD are hindered by the fermion sign problem, a fundamental technical obstacle inherent to any path integral representation of Fermi systems at finite density \cite{Philipsen:2012nu}. By breaking the balance between baryonic matter and anti-matter at high temperatures in QCD, the crossover is expected to end at a critical end point (CEP) and then evolve into a first-order phase transition. The CEP is characterized by the divergence of net-baryon number fluctuations. Understanding the emergence of critical phenomena in the theory of strong interactions has become a cardinal challenge not only for theory but also for experiments.  Depending on the location of the CEP in the temperature ($T$) and baryon chemical potential ($\mu_B$) axes of the QCD phase diagram, its effects may be probed using heavy ion collisions \cite{Stephanov:1998dy}. An experimentally-driven search for the QCD critical point is possible \cite{Adamczyk:2013dal} by systematically decreasing the center-of-mass energy/per nucleon ($\sqrt{s}$) of colliding ion beams, which enhances the amount of matter over anti-matter produced in these reactions. The first phase of such a beam energy scan (BES) program took place at RHIC and future runs with increased luminosity are scheduled for 2019-2020 after an upgrade of the machine. Fixed target experiments, reaching even larger baryon densities, will become fully operational in the near future \cite{Meehan:2017cum,Ablyazimov:2017guv}. 

In the absence of first principle lattice results in the baryon rich regime of QCD, effective approaches must be used to guide the experimental search for the critical point in heavy ion collisions. To be deemed realistic, such models must meet the following necessary requirements. First, the effective approaches must reproduce the thermodynamics of QCD in the crossover region at zero baryon density, as determined by lattice QCD. The other (more stringent) requirement is that for the temperatures probed in heavy ion collisions the system behaves as nearly perfect liquid. In this work we show that a model constructed using the holographic correspondence \cite{Maldacena:1997re}, a well-known tool developed in string theory, fulfills these requirements allowing one to determine the properties of the hot and baryon rich QGP liquid with unprecedented precision.

\section{Results}
\label{sec:results}

Through the holographic correspondence, calculations in strongly coupled non-Abelian gauge theories in four space-time dimensions at finite temperature and density can be performed using black hole solutions of classical theories of gravity in higher space-time dimensions. This approach has been previously applied to study some important aspects of the strongly coupled quark-gluon plasma \cite{CasalderreySolana:2011us} and also a variety of problems in condensed matter physics \cite{Zaanen:2015oix}. One of its main successes is the explicit derivation of nearly perfect fluid behavior at strong coupling, quantified by $\eta/s =1/4\pi$ \cite{Kovtun:2004de} (in natural units where $c=\hbar=k_B=1$), which is broadly compatible with recently extracted bounds for this quantity in heavy ion collisions \cite{Bernhard:2016tnd}.

In the holographic approach used in this work, conformal invariance in the plasma is dynamically broken by a real scalar field in the bulk, which roughly takes into account effects from the QCD running coupling, and an additional $U(1)$ gauge field is introduced in the dual gravity model to simulate the baryon charge and its corresponding chemical potential $\mu_B$. A similar approach was used in \cite{DeWolfe:2010he}, but contrary to that case, our construction provides a self-consistent gravitational setup with no extra free parameters besides the ones already featured in the gravity action. We numerically solve the corresponding five dimensional holographic equations of motion for the metric, the scalar field, and the gauge field to construct over two million charged black hole solutions (see the appendix), each one of them corresponding to a point in the $T-\mu_B$ phase diagram of the dual strongly coupled plasma. The parameters of the dual gravitational theory are fixed in order to reproduce two crucial quantities obtained through lattice simulations of QCD with 2+1 flavors with physical quark masses at zero baryon density: the entropy density \cite{Borsanyi:2013bia} and the second-order baryon susceptibility \cite{Bellwied:2015lba} $\chi_2$, which measures the equilibrium response of the baryonic density to a change in the chemical potential.  After imposing these constraints at zero baryon density, the model correctly predicts many other thermodynamic quantities compared to Lattice QCD at $\mu_B=0$. Additionally, predictions can also be made for the behavior of thermodynamic and transport quantities at finite $\mu_B$. This procedure, which we call \emph{black hole engineering}, is uniquely suited to investigate the baryon rich regime of QCD since it not only quantitatively reproduces the relevant results from the theory of strong interactions at finite temperature, but it also naturally incorporates the nearly perfect fluid property of the plasma (see the appendix).

\begin{figure*}
	\begin{center} 
		\includegraphics[width=0.8\textwidth]{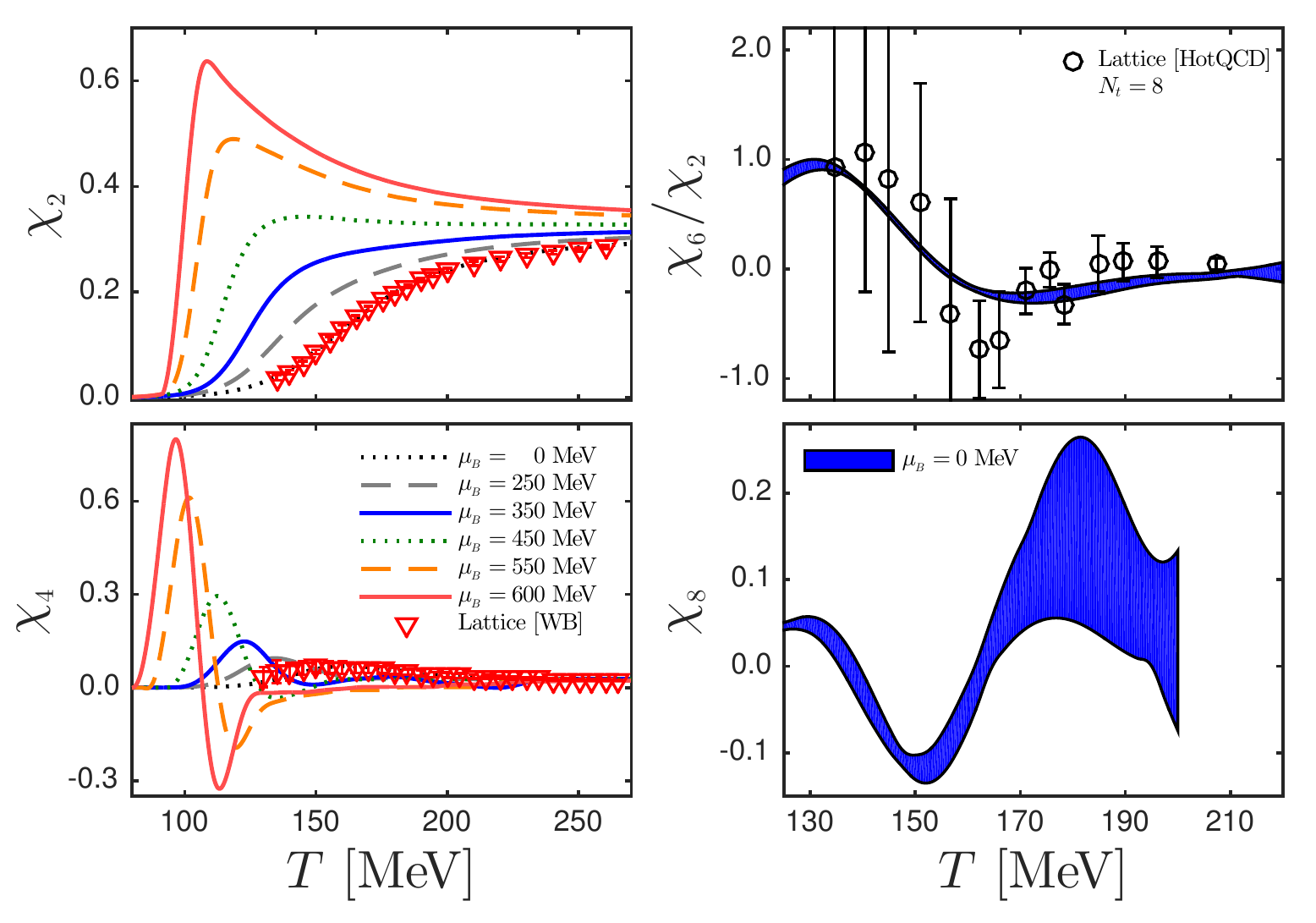}
	\end{center} 
	\caption{\label{fig1}(Color online) Baryon number susceptibilities ($\chi_n$) as functions of the temperature ($T$) for different values of the baryon chemical potential ($\mu_B$) computed using holographic black hole engineering. $\chi_2$ (upper left panel) and $\chi_4$ (lower left panel) are shown for values of the chemical potential in a range between 0 and 600 MeV; $\chi_6/\chi_2$ (upper right panel) and $\chi_8$ (lower right panel) are shown at $\mu_B=0$. The lattice results are from Refs.\ \cite{Bazavov:2017dus,Bellwied:2015lba}. We remind the reader that, while $\chi_2(T)$ at $\mu_B=0$ is used to fix the parameters of the holographic framework, all other quantities are predictions of our approach. The error-band on our predictions for $\chi_6/\chi_2$ and $\chi_8$ denotes the uncertainty in the numerical calculation of the higher order derivatives (see Methods).}
\end{figure*}

In the vicinity of the critical point, the higher order baryon number susceptibilities defined as $\chi_n(T,\mu_B)=\partial^n(P/T^4)/\partial(\mu_B/T)^n$, where $P=P(T,\mu_B)$ is the pressure of the system, diverge with different powers of the correlation length $\xi$ \cite{Stephanov:2008qz}. To investigate the onset of critical behavior, after determining the pressure via holography, numerical derivatives are taken to determine the second, fourth, sixth, and eight order baryon number susceptibilities shown in Fig.\ \ref{fig1}. One can see that $\chi_2(T,\mu_B)$ begins to develop a peak for large chemical potentials, which will then evolve into a divergence at the critical point. The figure also shows the available lattice results for $\chi_2$, $\chi_4$ \cite{Bellwied:2015lba} and $\chi_6/\chi_2$ \cite{Bazavov:2017dus} as a function of $T$. Our predictions for $\chi_4(T)$ and $\chi_6(T)/\chi_2(T)$ have a remarkable agreement with lattice QCD results. As for $\chi_8(T)$, our prediction exhibits the features expected from universality arguments \cite{Karsch:2009zz} and can be readily compared to lattice QCD results once they become available. 

\begin{figure*}
\begin{center}
	\begin{tabular}{cc}
		\includegraphics[width=0.45\textwidth]{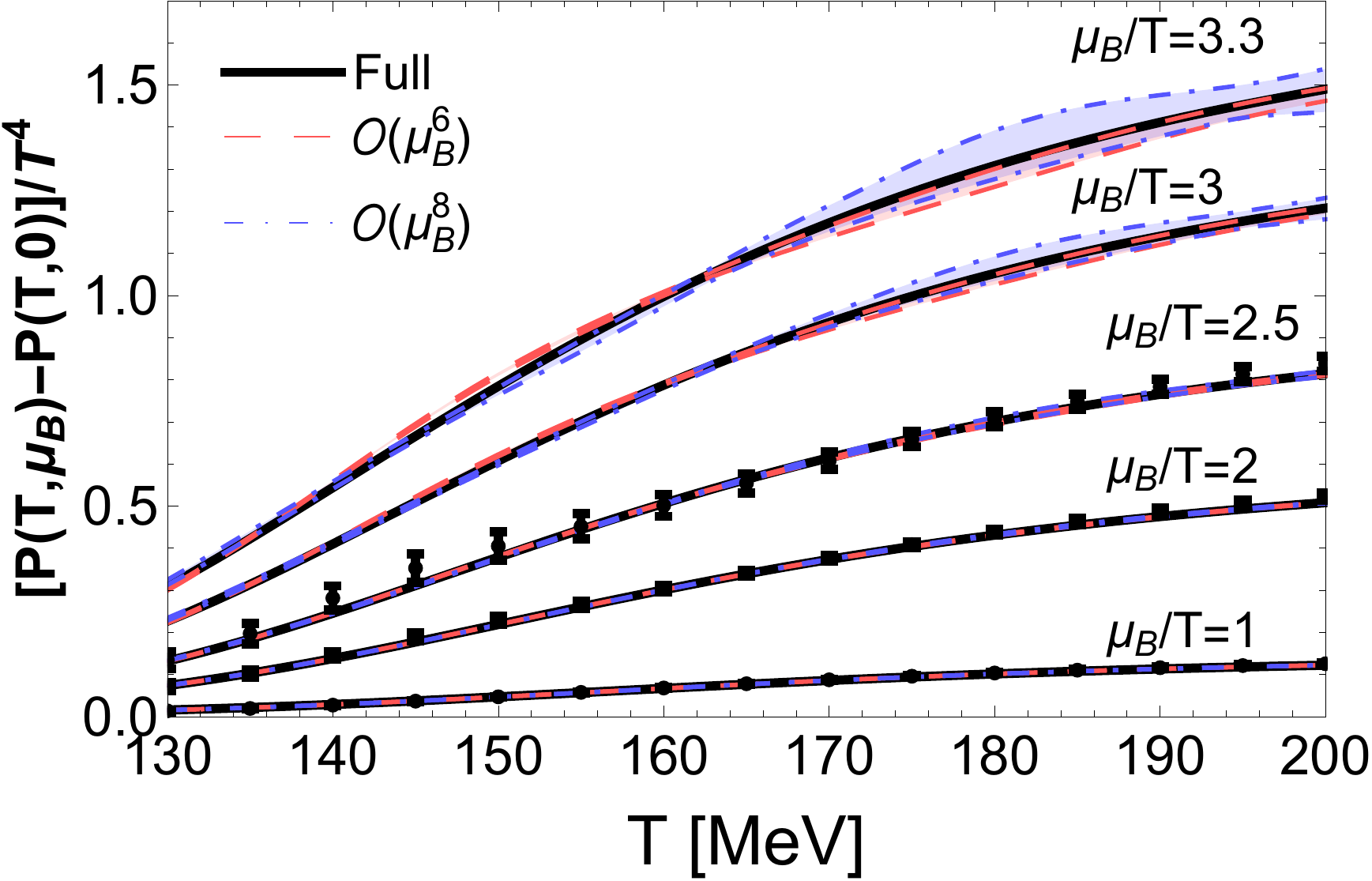} & 
		\includegraphics[width=0.45\textwidth]{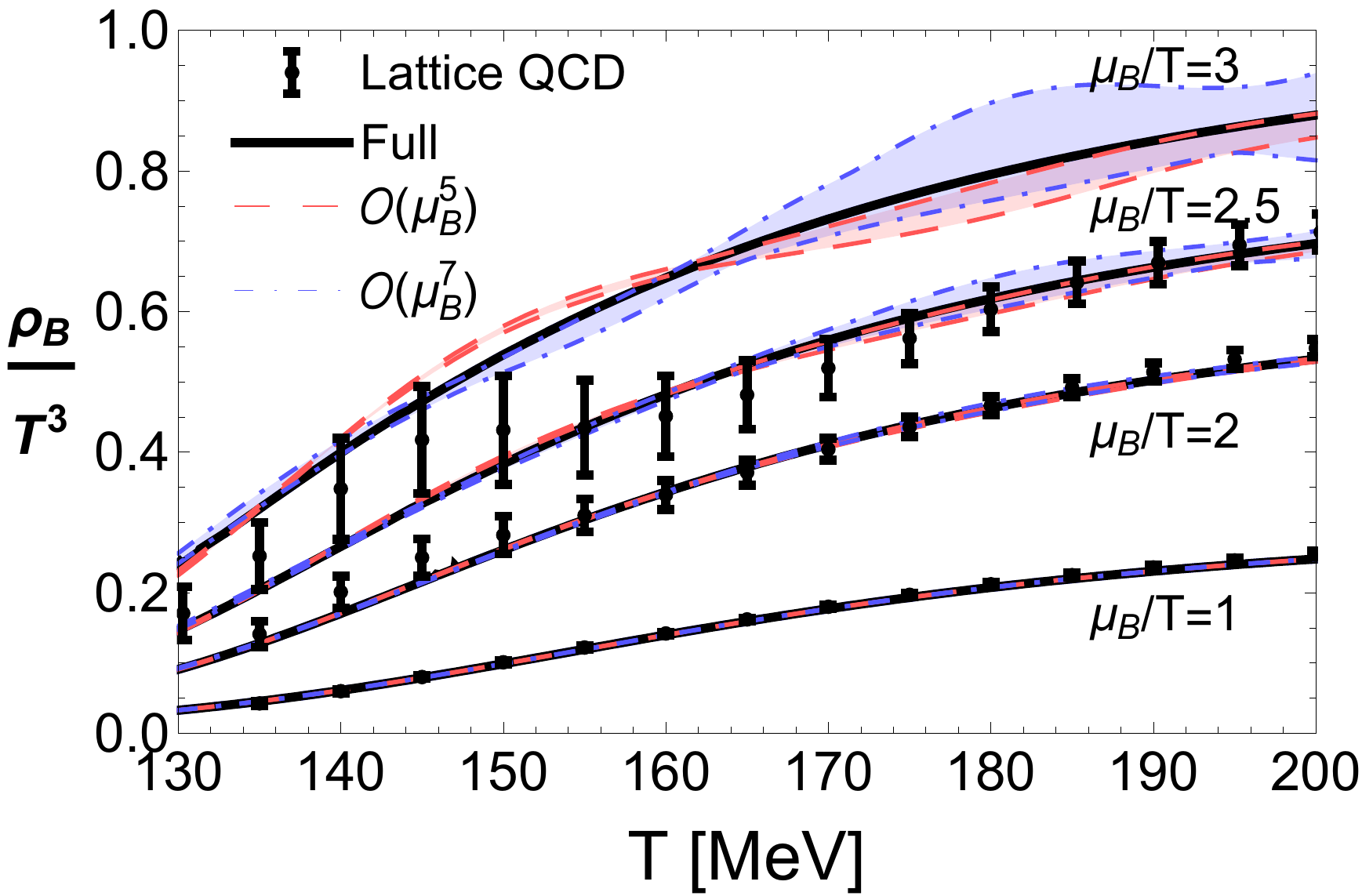}
	\end{tabular}
	\end{center}
\caption{\label{recon_EOS}(Color online) The $\mu_B$-dependent contribution to the pressure (left) and the baryon density (right) as functions of $T$ for different values of $\mu_B/T$. The solid curves correspond to the full holographic result computed using black hole engineering. The bands denote the holographic results reconstructed through a power series expansion up to different orders in $\mu_B/T$, using the quantities displayed in Fig.\ \ref{fig1}. The points correspond to the reconstructed Taylor series up to $\mathcal{O}(\mu_B^6)$ for the pressure and $\mathcal{O}(\mu_B^5)$ for $\rho_B$ computed on the lattice \cite{Bazavov:2017dus}.}
\end{figure*}

Using the higher order susceptibilities one may reconstruct the system's pressure and baryon density $\rho_B=\chi_1 T^3$ as a Taylor series in powers of $\mu_B/T$ as follows
\begin{align}
\label{eqn:taylor}
&\frac{P(T,\mu_B)-P(T,\mu_B=0)}{T^4}=\sum_{n=1}^\infty\frac{1}{(2n!)}\chi_{2n}(T)\left(\frac{\mu_B}{T}\right)^{2n},\\ 
\label{rhotaylor}
&\frac{\rho_B(T,\mu_B)}{T^3}=\sum_{n=1}^\infty\frac{1}{(2n-1)!}\chi_{2n}(T)\left(\frac{\mu_B}{T}\right)^{2n-1}.
\end{align}
In Fig.\ \ref{recon_EOS} the pressure difference in \eqref{eqn:taylor}, calculated in the holographic model with no truncations, is compared to the lattice QCD results from Ref.\ \cite{Bazavov:2017dus}. Additionally, the reconstructed holographic pressure truncated at order $\mathcal{O}(\mu_B^6)$ and $\mathcal{O}(\mu_B^8)$ is also shown (the bands reflect the numerical uncertainties in the calculations of $\chi_6(T)$ and $\chi_8(T)$, see Methods). Our analysis not only confirms the applicability of the $\mathcal{O}(\mu_B^6)$ truncation done in \cite{Bazavov:2017dus} for $\mu_B/T \leq 2$ but it also predicts that the inclusion of $\chi_8(T)$ into the expansion extends the domain of applicability of the Taylor series to at least $\mu_B/T\sim 2.5$ (further discussion can be found in the appendix).



\begin{figure*}[tb]
	\begin{center} 
	\begin{tabular}{cc}
		\includegraphics[width=0.5\textwidth]{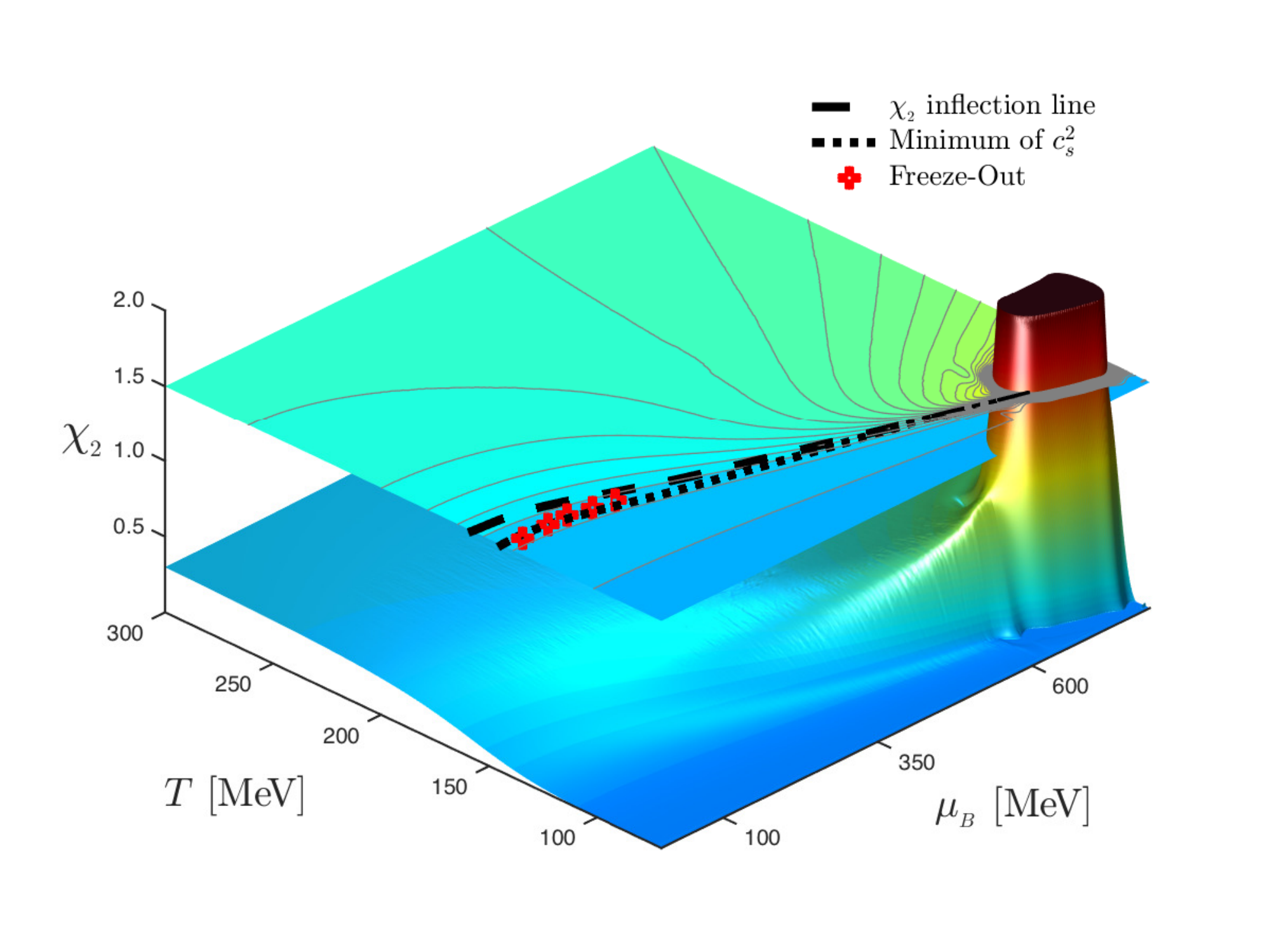} &
		\includegraphics[width=0.4\textwidth]{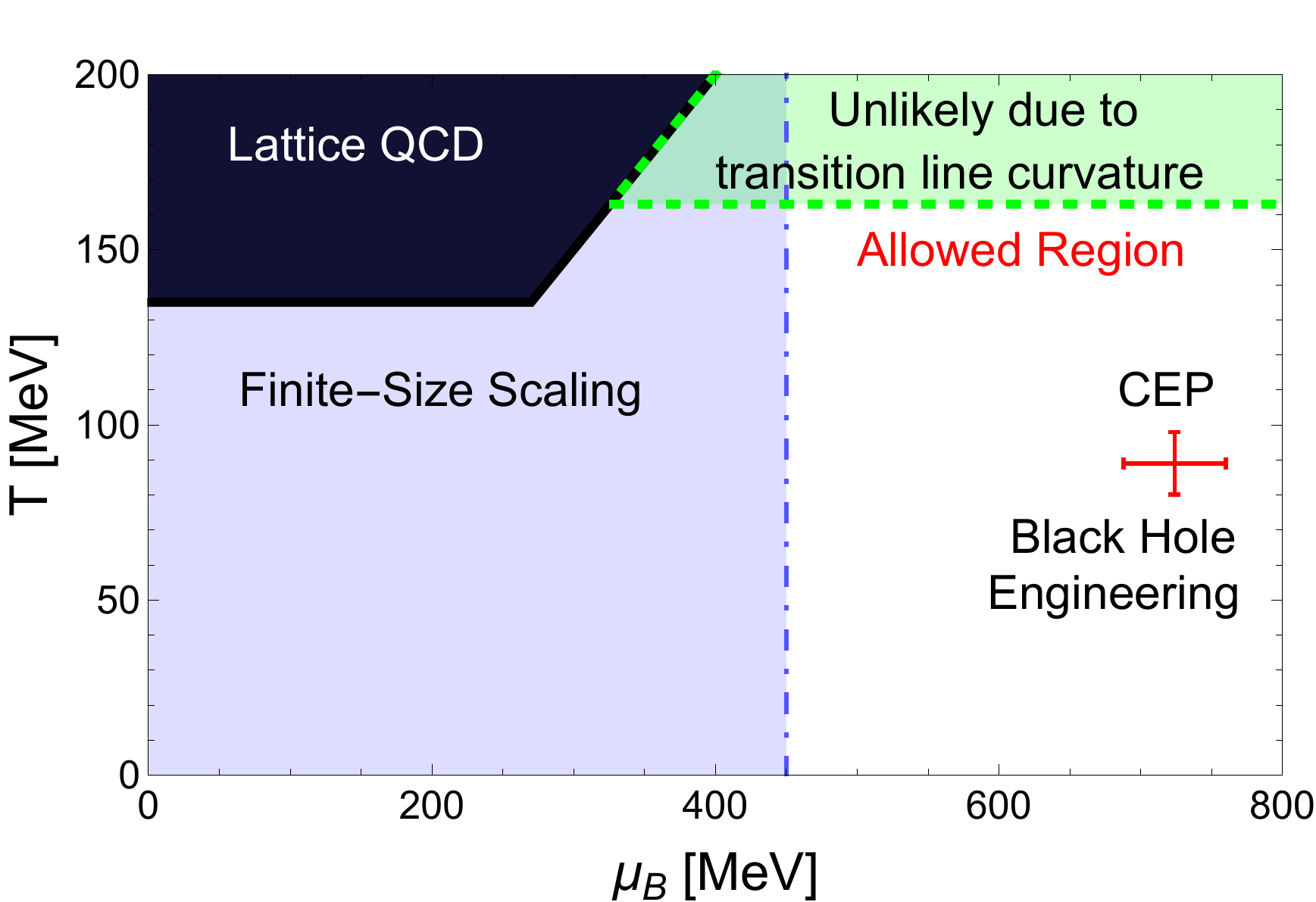}
		\end{tabular}
	\end{center} 
	\caption{\label{fig3}(Color online) The left plot shows the behavior of the baryon susceptibility $\chi_2$ in the $T-\mu_B$ plane determined from black hole engineering. As the chemical potential increases $\chi_2(T,\mu_B)$ develops a peak, which turns into a divergence at the critical point located at $T_{CEP}=89$ MeV and $\mu_B^{CEP}=724$ MeV (see the appendix). The upper plane in the left plot shows the phase diagram obtained through our method, with our chemical freeze-out points in red. The dashed line corresponds to the location of the inflection point of $\chi_2$ in the $T-\mu_B$ plane, one of the quantities chosen to characterize the phase transition line. The dotted line gives the location of the minimum of the speed of sound squared, $c_s^2$, in the phase diagram. The right panel shows the regions in the QCD phase diagram where the presence of a critical point has been excluded  by current lattice QCD constraints \cite{Bazavov:2017dus} and a finite-size scaling analysis \cite{Fraga:2011hi}. Temperatures above 155 MeV are also unlikely due to constraints from the curvature of transition lines \cite{Bellwied:2015rza}. The location of the critical point in the phase diagram that we found in this work, taking into account our systematic uncertainties (see the appendix), is also shown.}
\end{figure*}

By carefully inspecting the behavior of $\chi_2$ and $\rho_B$, using the best set of parameters for the holographic model (see the appendix), we find a critical point in the phase diagram at $T_{CEP}=89$ MeV and $\mu_B^{CEP}=724$ MeV, which should be compared to the original holographic study in Ref.\ \cite{DeWolfe:2010he} that found $(T_{CEP},\mu_B^{CEP})=(143,783)$ MeV using previous lattice results for which the transition temperature was $\sim 190$ MeV, instead of the more precise current value $\simeq155$ MeV \cite{Borsanyi:2010bp}. A more detailed investigation of the effects from uncertainties in the lattice results used in the determination of the parameters of our model (see the appendix) shows that $T_{CEP}$ may change by at most $13\%$ and $\mu_B^{CEP}$ by at most $5\%$. Fig.\ \ref{fig3} (left) shows $\chi_2$ in the $T-\mu_B$ plane and its rapid increase near the critical point. The critical point is located along the line $\mu_B/T \sim 8.1$ in the phase diagram, which is beyond the reach of current lattice QCD calculations where $\mu_B/T \lesssim 2$ \cite{Bazavov:2017dus}. We show in the appendix that the location of this critical point cannot be reliably obtained via an analysis of the radius of convergence of the series in powers of $\mu_B/T$ constructed using only the results for $\chi_{n}(T)$ with $n=2,4,6,8$.

Since the transition is a smooth crossover at small values of $\mu_B$, there is no unique definition of the transition temperature. In practice, this quantity is usually identified with the inflection point (steepest rise) or maximum/minimum of some quantity which would be sensitive to a change in the relevant degrees of freedom of the system from hadrons to quarks and gluons \cite{Bellwied:2015lba}. Here we choose two such quantities: the inflection point of $\chi_2$ and the minimum in the speed of sound squared $c_s^2$ (see Methods). The phase transition lines thus obtained are shown in the upper plane of the left plot in Fig.\ \ref{fig3}, together with the contour lines for $\chi_2$. Even though these quantities define different transition temperatures at $\mu_B=0$, they converge at the critical point, as expected. Finally, the right panel of Fig.\ \ref{fig3} shows our critical point (including systematic uncertainties) and the regions of the QCD phase diagram where the presence of a critical point has been already excluded using different approaches \cite{Bazavov:2017dus,Fraga:2011hi}. Regions where $T> 155$ MeV are also unlikely to display a CEP due to the known behavior of the curvature of transition lines at low $\mu_B$ \cite{Bellwied:2015rza}.

In the following we discuss the consequences of our results to the ongoing experimental search for the QCD critical point using heavy-ion collisions. We begin by providing our estimate for the heavy-ion collision center-of-mass energy that could probe the values of $T_{CEP}$ and $\mu_B^{CEP}$ found here. Experimentally measured mean particle yields in heavy-ion collisions have long been used, in the context of statistical hadronization models (SHM) \cite{BraunMunzinger:2007zz}, to extract the dependence of the pair $(T,\mu_B)$ of the matter created with the collision energy $\sqrt{s}$ at the point where  hadrons reach chemical equilibrium (i.e., chemical freeze-out) \cite{Andronic:2008gu}. Another way to estimate this $\sqrt{s}$ dependence comes from the measurement of moments of the measured net-particle distributions. In fact, the mean over the variance of the distribution is equivalent to the ratio of susceptibilities $\chi_1/\chi_2$ and a comparison between theory and experiment for this and other similar ratios may also be used to determine how $(T,\mu_B)$ varies with $\sqrt{s}$ \cite{Karsch:2010ck,Gupta:2011wh}. Both methods were used here to gauge the uncertainties in such a mapping and the details of this analysis can be found in the Methods section. The chemical freeze-out points, displayed in red in the upper plane of Fig.\ \ref{fig3}, were extracted through a comparison of holographically computed baryon number susceptibilities and experimental data for net-proton fluctuations from \cite{Adamczyk:2013dal} (see Fig.\ \ref{fig6} in the Methods section) and they were found to lie along the transition line defined by the minimum of $c_s^2$ when $\sqrt{s}\geq 27$ GeV. By consistently extrapolating this behavior towards smaller collision energies, taking into account different sources of systematic uncertainties, we find that the critical point of the model could be probed using heavy ion experiments with center-of-mass energy in the range $\sqrt{s}=2.5-4.1$ GeV (see Methods). These collision energies are below the current plans for the 2nd phase of the RHIC BES operating in collider mode (where the minimum is $\sqrt{s}=7.7$ GeV) but they are within the reach of the HADES experiment \cite{Agakishiev:2015bwu}, the planned Fixed Target (FXT) program also at RHIC \cite{Meehan:2017cum}, and the future Compressed Baryonic Matter (CBM) experiment at FAIR \cite{Ablyazimov:2017guv}.
 

\begin{figure}[ht!]
\begin{center}
		\includegraphics[width=0.45\textwidth]{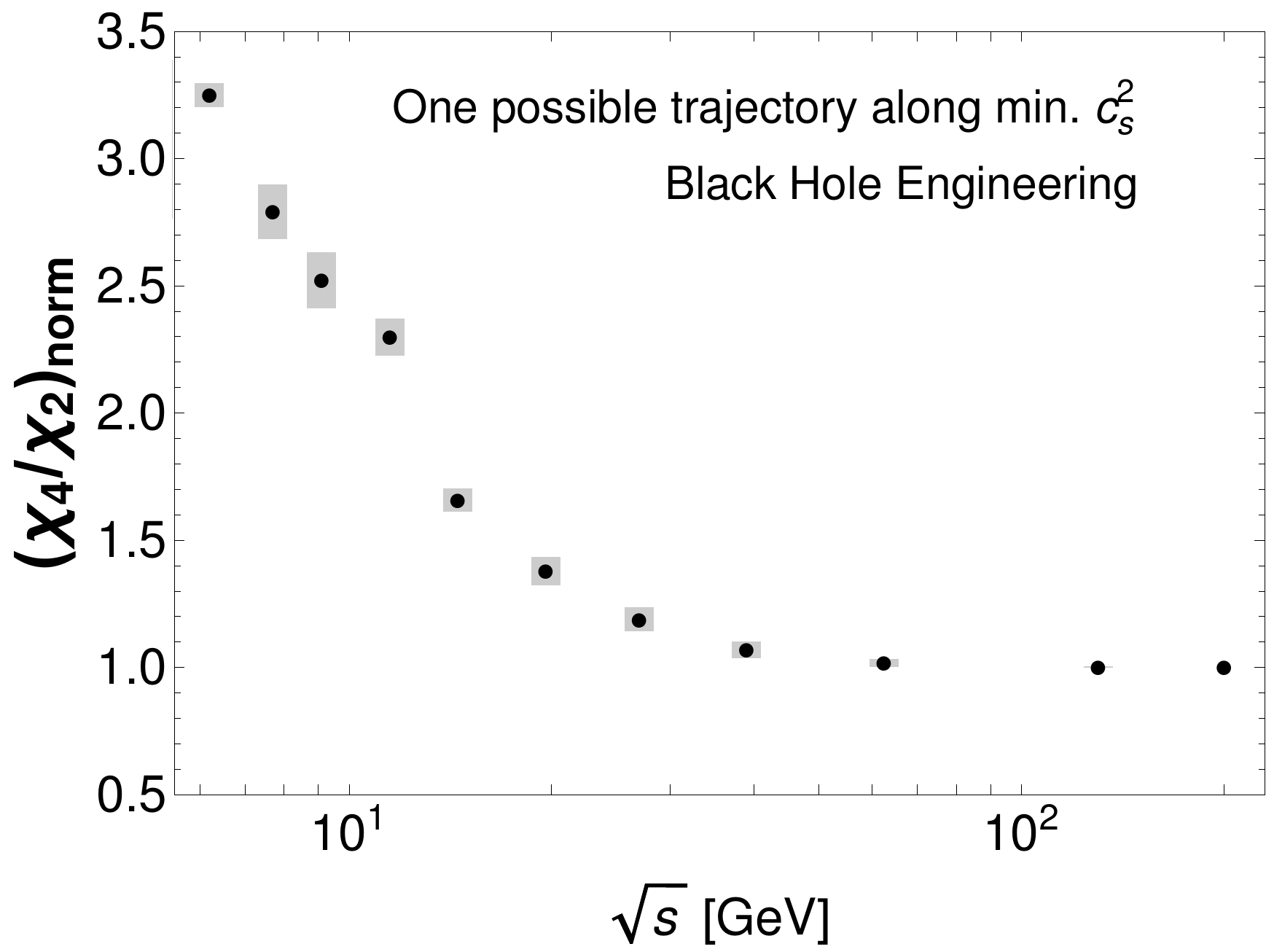}  
	\end{center}
\caption{\label{fig4}(Color online) Ratio between baryon number susceptibilities, $\chi_4/\chi_2$, for $0-5\%$ most central Au-Au collisions (normalized by its value at $\sqrt{s}=200$ GeV) as a function of the center-of-mass collision energy $\sqrt{s}$ computed via black hole engineering. The points are computed along the transition line given by the minimum of $c_s^2$ (with error bars described in Methods), which agrees with the chemical freeze-out line extracted from the moments of net-proton fluctuations for $\sqrt{s}\geq 27$ GeV. The ratio $\chi_4/\chi_2$ considerably increases along this particular trajectory, even though the values of $\sqrt{s}$ involved are not in the region corresponding to the critical point of the model ($\sqrt{s}=2.5-4.1$ GeV). We note that this is only one possible trajectory where the freeze-out line passes through the CEP. Alternatively, the freeze-out line could pass below the CEP, which would change the behavior of $\chi_4/\chi_2$ with $\sqrt{s}$.}
\end{figure}
 
While universality arguments dictate the sign of $\chi_4/\chi_2$ sufficiently close to the critical region \cite{Stephanov:2011pb}, if the QCD critical point follows our prediction and is located at moderately large values of $\mu_B$ and low temperatures, depending on the trajectory followed in the phase diagram this ratio can already display some nontrivial behavior for values of $\sqrt{s}$ larger than those associated with the critical point. Using the aforementioned methods to establish how $T$ and $\mu_B$ depend on $\sqrt{s}$ at freeze-out, and our results for the susceptibilities across the phase diagram, we show a calculation for the ratio $\chi_4/\chi_2$ normalized by its value at $\sqrt{s}=200$ GeV (to minimize its dependence on corrections from particle decays, acceptance cuts, and the fact that experimentally only net protons are measured) in Fig.\ \ref{fig4} for the values of $\sqrt{s}$ within the RHIC BES. The points are computed on the transition line, $T=T(\mu_B)$, defined by the minimum of $c_s^2$ (with error bars described in detail in the Methods section). As mentioned above, this transition line agrees with the chemical freeze-out points extracted here from the moments of net-proton fluctuations for $\sqrt{s}\geq 27$ GeV. One can see that $\chi_4/\chi_2$ monotonically increases in this case even though one is still outside the critical region located at $\sqrt{s}=2.5-4.1$ GeV. By using other choices for the chemical freeze-out line at low $\sqrt{s}$ (still outside the critical region), non-monotonic behavior for $\chi_4/\chi_2$ with $\sqrt{s}$ can be found that does not follow from the universality arguments of \cite{Stephanov:2011pb}. This should be kept in mind when comparing model calculations to upcoming experimental data from RHIC and other facilities.

Overall, the main result of our analysis is the prediction of the existence and location of a critical point on the phase diagram of QCD, situated in the allowed region of the phase diagram in Fig.\ \ref{fig3} (right), which may be investigated by the next generation of heavy ion experiments designed to probe the novel properties of the hot and baryon rich quark-gluon plasma using center-of-mass collision energies in the range $\sqrt{s}=2.5-4.1$ GeV.

\section{Methods}
\label{sec:methods}

\subsection*{Numerical calculation of higher order baryon number susceptibilities}

The higher order baryon susceptibilities may also be computed through the derivatives of the baryon density, which is proportional  to the first baryonic susceptibility (\(\chi_1\)), with respect to $\mu_B/T$ for fixed $T$. The baryon density is calculated using $N=2 \times 10^{6}$ holographic black hole solutions (see the appendix). The original data set for $\chi_1$ is not equally spaced in the (\(T,\mu_B\)) plane and an additional procedure has to be used to determine $\chi_1$ on a regular grid. This is done by interpolating $\chi_1$ and then computing its value on an equally spaced grid.  The high precision derivatives themselves are calculated within a smaller range of temperatures and chemical potentials in the interval \(T=[65-450]\) MeV and \(\mu_B=[0-600]\) MeV.  A master grid is created in the (\(T,\mu_B\)) plane, which is divided into square nodes of width \(\Delta{T}=5\) MeV and \(\Delta{\mu_B}=20\) MeV. Each node is individually interpolated using the points inside the node and its neighbor nodes using thin-plate splines. The thin-plate splines interpolation was chosen over nearest neighbor, polynomial, cubic spline, and bi-harmonic interpolations because it provided the best surface interpolation for the baryonic susceptibilities. The neighbor node points are used to eliminate boundary effects in the interpolation. On the master grid, we create extra nodes outside its boundary and impose several constraints. For the \(\mu_B=0\) axis we reflect the points depending on the symmetry of the given susceptibility (even (odd) susceptibilities have even (odd) parity when reflected along the \(\mu_B=0\) axis). For the \(T=65\) MeV axis, the extra nodes are set to zero, while for the other two axes, (\(\mu_B=600\) MeV and \(T=450\) MeV), the nodes are set to have a constant derivative equal to the value of the one at the corresponding boundary of the master grid. Using this interpolation scheme,  \(\chi_1\) is obtained via the master grid using an equally spaced grid of points with separation 0.25 MeV in \(T\) and \(\mu_B\).

The next order susceptibility, in this case \(\chi_2\), is also obtained from the interpolation scheme; however this susceptibility, which is the derivative with respect to \(\mu_B\) of the interpolated points for $\chi_1$, contains noise associated with the interpolation. The noise makes it impossible to calculate the next derivative ($\chi_3$) starting from this raw data set for $\chi_2$ and a filtering procedure must be employed. In this paper the noise is removed by using a Savitzky-Golay (SG) filter, a low-pass filter well adapted for smoothing out noisy data.  Once the filter has been applied to the signal, a smooth $\chi_2$ is available to seed the master grid, which will then repeat the procedure to calculate the next susceptibility.

The SG-filter preserves the original shape and features of the signal better than other common types of filters. This method performs a least squares fit of the \(N_T\) and \(N_{\mu_B}\) number of neighbors of each data point to a polynomial of degree \(k\) and takes the calculated central point of the fitted polynomial curve as the new smoothed data point. The baryonic susceptibilities have a well-defined structure without any abrupt changes and, for that reason, we choose the input parameter \(k=3\). This polynomial allows us to remove rapidly varying structures that are created  by numerical noise. On the other hand, the input parameters \(N_T\), \(N_{\mu_B}\) are chosen according to the degree of non-smoothness of each susceptibility, using values that are as small as possible to avoid the generation of numerical artifacts.  

We varied the input parameters of the SG-filter to test the robustness of our numerical procedure for the calculation of the higher order susceptibilities. We verified that the results are robust enough to determine the $T$ and $\mu_B$ dependence of the susceptibilities up to $\chi_4$. In fact, the behavior of the susceptibilities in the $(T,\mu_B)$ plane does not change for a large number of input parameters though an attenuation of the peaks found in the susceptibilities occurs for very strong filters, especially at large \(\mu_B\). Therefore, we only consider filter parameters that do not change the peaks of the susceptibilities by more than \(5\%\). On the other hand, the baryonic susceptibilities  \(\chi_6(T)\) and \(\chi_8(T)\) at \(\mu_B=0\) were computed directly from \(\chi_4(T,\mu_B)\) using finite differences. This calculation gives the results shown in Fig.\ \ref{fig1}, which possess an uncertainty band associated with the variation of the width of the finite difference procedure and the effects coming from varying the SG-filter parameters.

\subsection*{Details about the chemical freeze-out analysis}

In relativistic heavy ion collisions it is generally assumed that the particle yields are fixed at chemical freeze-out and, therefore, information on the chemical equilibrium temperature(s) and $\mu_B$ can be extracted by comparing particle yields computed using theoretical models to experimental data.  Statistical hadronization models allow one to calculate such particle yields using a relatively simple framework called the hadron resonance gas (HRG) model, where hadrons are assumed to be non-interacting point-like particles. This type of model has been quite successful for many years \cite{BraunMunzinger:2007zz} and expressions for $T$ and $\mu_B$ as functions of the collision energy $\sqrt{s}$ are well-known \cite{Andronic:2008gu}. However, these purely hadronic models contain no information about the QGP phase nor any possible effects from critical phenomena, so one would expect that eventual difficulties in describing experimental data could appear at large enough baryon densities (close to the critical point).  

More recently, new experimental observables have been devised that focus on the event-by-event fluctuations of conserved charges \cite{Karsch:2010ck,Gupta:2011wh}.  For instance, by measuring the distribution of net-protons one can obtain a reasonable proxy for the distribution of net-baryons. Then, the moments of this distribution may be directly compared to first principle lattice QCD calculations to extract the freeze-out line \cite{Bazavov:2012vg,Borsanyi:2013hza}. In Ref.\ \cite{Borsanyi:2014ewa}, at a set energy $\sqrt{s}$ the $M/\sigma^2$ (mean over the variance) of the net-proton distribution and the $M/\sigma^2$ of the net-electric charge distribution (which includes pions, protons, and kaons) are compared to lattice QCD results for the baryonic $\chi_1/\chi_2(T,\mu_B)$  and electric charge $\chi_1/\chi_2(T,\mu_B)$, respectively.  Then, one has two equations and two unknowns and can extract the corresponding $\left(T,\mu_B\right)$ pair at a specific $\sqrt{s}$, which gives the chemical freeze-out line. Note that ratios are always used to form volume-independent quantities. In the low $\mu_B$ region, Ref.\ \cite{Borsanyi:2013hza} finds a good description for the extracted $\left(T,\mu_B\right)$ between the hadron resonance gas model and lattice QCD.

\begin{figure*}
	\begin{center}
		\includegraphics[width=35pc]{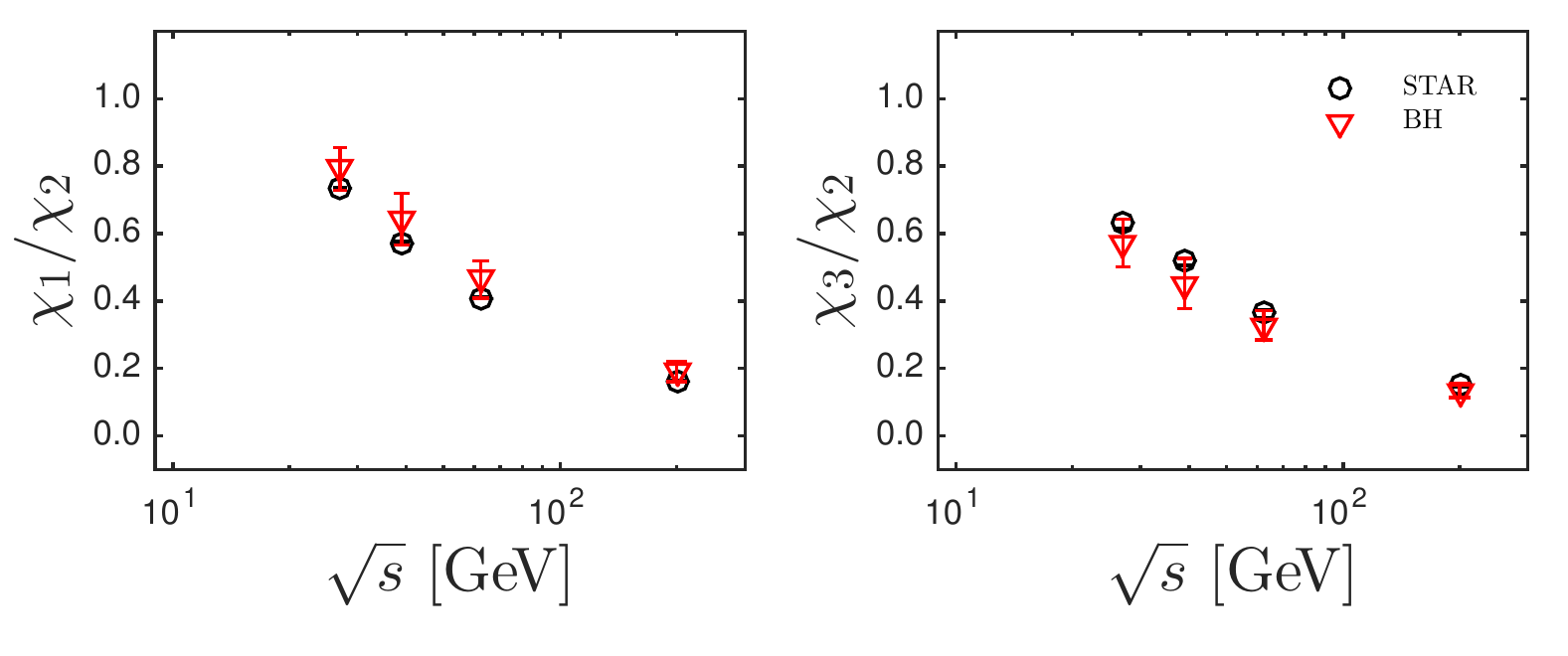}
	\end{center}
    \caption{\label{Fig:chisSTAR}(Color online) Collision energy dependence of $\chi_1/\chi_2$ and $\chi_3/\chi_2$ along the chemical freeze-out line computed using our black hole engineering model (red triangles) compared to the 0-5\% net-proton distribution Au-Au data from the STAR experiment \cite{Adamczyk:2013dal} for $\sqrt{s}\geq 27$ GeV. }
    \label{fig6}
\end{figure*}

As mentioned in the main text, the higher-order susceptibilities are more sensitive to criticality and, thus, one would not expect the hadron resonance gas model (or any other model involving only hadronic degrees of freedom) to adequately describe higher order susceptibilities as one approaches the critical region. Generally, the analysis based on susceptibilities produces a slightly lower temperature than SHM calculations \cite{Alba:2014eba}. Within our own model we only have one conserved charge -baryon number- so we can use $\chi_1/\chi_2(T,\mu_B)$ and $\chi_3/\chi_2(T,\mu_B)$ to extract $\left(T,\mu_B\right)$ as functions of $\sqrt{s}$ comparing to experimental data of $M/\sigma^2$ and $S\sigma$ (skewness times standard deviation) of the net-proton distribution. We compare our results for $\chi_1/\chi_2$ at freeze-out to the mean over the variance ($M/\sigma^2$) of net-protons and $\chi_3/\chi_2$ to the skewness times the variance ($\mathcal{S}\sigma$) of net-protons measured by STAR \cite{Adamczyk:2013dal} in Fig.\ \ref{Fig:chisSTAR}. This could not be done in the hadron resonance gas model where $\chi_1/\chi_2\sim \chi_3/\chi_2\sim 1$ and, in fact, hadronic models are known to miss the $\sqrt{s}$ dependence of higher order susceptibilities \cite{Adamczyk:2013dal}. One can see that our results can be reasonably matched to STAR data.  

When comparing $\chi_1/\chi_2(T,\mu_B)$ to $M/\sigma^2$ and $\chi_3/\chi_2(T,\mu_B)$ to $S\sigma$, one produces two different bands in $\left(T,\mu_B\right)$ after the inclusion of the experimental error. We then look for the point where either the bands overlap (or their nearest point) to extract the corresponding freeze-out pair $\left(T,\mu_B\right)$ at a certain $\sqrt{s}$ and our error bars are extracted from the width of the two bands at that point.  We remark that we are aware of the limitations of our model, which does not include strangeness or electric charge chemical potentials, or the acceptance cuts which match the experimental setup. For these reasons, when extracting the chemical freeze-out points we limit our analysis to the large collision energies $\sqrt{s}\geq 27$ GeV where such effects are expected to be small. In the end, we find chemical freeze-out temperatures and chemical potentials which are compatible to the ones obtained from the analysis of fluctuations in the HRG model \cite{Alba:2014eba} and lattice QCD \cite{Borsanyi:2013hza}. 

\begin{figure*}
	\begin{center}
		\includegraphics[width=35pc]{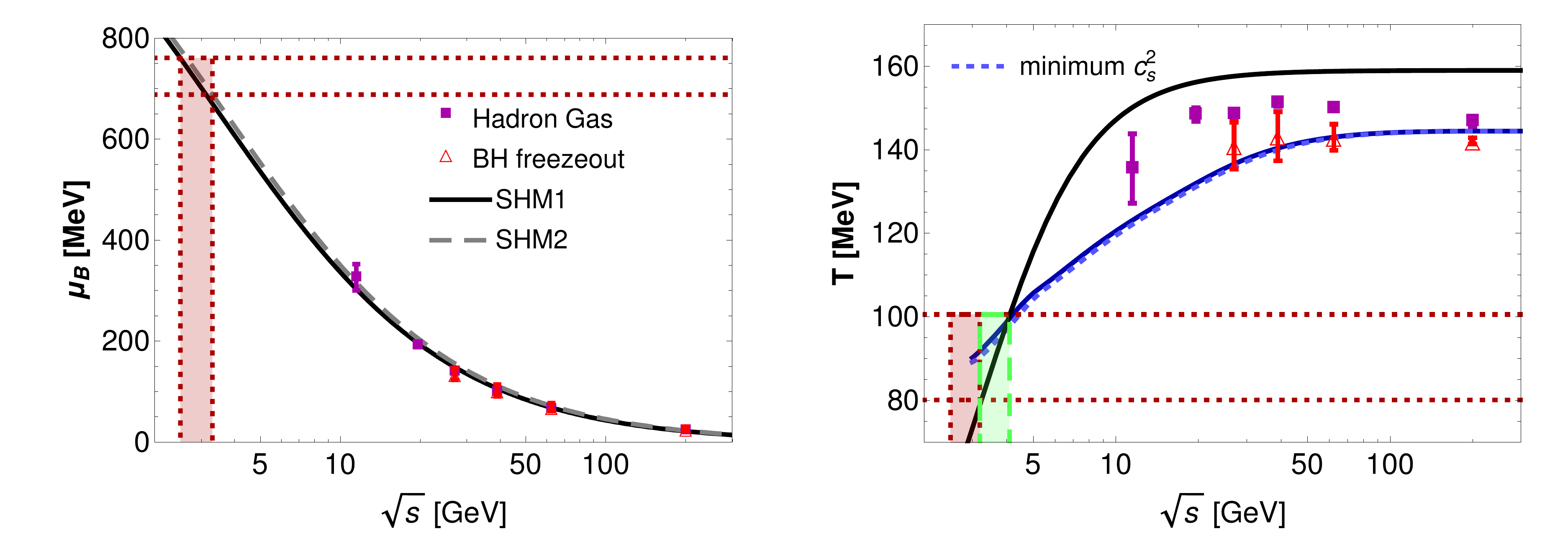}
	\end{center}
    \caption{\label{snn}(Color online). Collision energy dependence of the baryon chemical potential (left) and  temperature (right) at chemical freeze-out. The solid black line denotes the freeze-out line for $\mu_B(\sqrt{s})$ and $T(\sqrt{s})$  defined using the statistical hadronization model calculations of  \cite{Andronic:2008gu} (SHM1). The dashed grey curve corresponds to the parametrization for $\mu_B(\sqrt{s})$ from \cite{Cleymans:2005xv} (SHM2). The solid and dashed blue lines illustrate how the trajectory in the phase diagram that follows the minimum of $c_s^2$ changes with $\sqrt{s}$ using the parametrizations SHM1 and SHM2 for $\mu_B(\sqrt{s})$. The purple squares represent hadron resonance gas comparisons to net-proton and net-electric charge fluctuations from \cite{Alba:2014eba}. The red triangles represent the chemical freeze-out points extracted in this work, via a comparison between the $\chi_1/\chi_2$ and $\chi_3/\chi_2$ computed using black hole engineering and the corresponding net-proton experimental data from STAR \cite{Adamczyk:2013dal} (see Fig.\ \ref{fig6}). The bands are used to find the value of $\sqrt{s}$ corresponding to $T_{CEP}$ and $\mu_B^{CEP}$ of the critical point including the combined effect from uncertainties coming from the parametrizations $T(\sqrt{s})$ and $\mu_B(\sqrt{s})$ and other sources from the holographic calculations.}
\end{figure*}

In Fig.\ \ref{snn} (left) we show $\mu_B(\sqrt{s})$ (purple squares) extracted using the susceptibilities calculated within the HRG model \cite{Alba:2014eba}; $T(\sqrt{s})$ for the same model is shown on the right panel. Our results using black hole engineering and $\chi_1/\chi_2$ and $\chi_3/\chi_2$  to extract $\mu_B$ as a function of $\sqrt{s}$ lead to the red triangles shown in Fig.\ \ref{snn}, which are compatible with the results from the statistical hadronization models \cite{Andronic:2008gu} (SHM1) and \cite{Cleymans:2005xv} (SHM2). Thus, in order to estimate $\mu_B(\sqrt{s})$ at lower energies we use the two parametrizations, SHM1 and SHM2, mentioned above. In Fig.\ \ref{snn} (right) $T(\sqrt{s})$ from SHM1 is shown in solid black. 

Another method to determine the phase transition from the QGP to the hadron gas phase involves looking at inflection points or extrema of thermodynamic quantities.  Thus, we also consider the minimum of $c_s^2$, which allows us to determine a different curve $T=T(\mu_B)$ in the phase diagram.  Using the two SHM parametrizations for $\mu_B(\sqrt{s})$ one obtains the solid and dashed blue $T(\sqrt{s})$ curves,  which are shown in Fig.\ \ref{snn} (right). We note that our freeze-out points for $\sqrt{s}\geq 27$ GeV lie along this $c_s^2$ transition line. In our calculations of the normalized ratio $\chi_4/\chi_2$ of net-baryon number shown in Fig.\ \ref{fig4} we included both the different transition lines defined by the minimum of $c_s^2$ and the inflection point of $\chi_2$ and also the difference between the two different SHM parametrizations for $\mu_B(\sqrt{s})$ into our error bars. Furthermore, in Fig.\ \ref{snn} one can find the vertical colored bands we used to estimate the values of $\sqrt{s}$ corresponding to $T_{CEP}$ and $\mu_B^{CEP}$ of the critical point, which include the combined effect from uncertainties coming from the parametrizations $T(\sqrt{s})$ and $\mu_B(\sqrt{s})$ and also the other sources of uncertainty associated with the holographic calculations discussed in the appendix. The latter generate the dashed  horizontal lines in Fig.\ \ref{snn} while the vertical colored bands are obtained by finding the values of $\sqrt{s}$ in both plots where these horizontal lines cross the $T(\sqrt{s})$ and $\mu_B(\sqrt{s})$ curves from statistical models and from our curve corresponding to the minimum of $c_s^2$. The final range for the values of $\sqrt{s}$ corresponding to the critical point region mentioned in the main text, $\sqrt{s}=2.5-4.1$ GeV, is obtained by combining the colored systematic uncertainty bands in Fig.\ \ref{snn} (right).

\begin{acknowledgments}
We thank R.~Bellwied, P.~Parotto, and K.~Meehan for helpful comments and S.~Sharma for providing the tables that contain the publicly available results of Ref.\ \cite{Bazavov:2017dus}. J.N., R.R., and R.C. thank S.~Finazzo for insightful discussions on the gauge/gravity duality at nonzero baryon density. J.N. thanks the Funda\c c\~ao de Amparo \`a Pesquisa do Estado de S\~ao Paulo (FAPESP) and Conselho Nacional de Desenvolvimento Cient\'ifico e Tecnol\'ogico (CNPq) for support. R.C. was supported by FAPESP grant 2016/09263-2. R.R. acknowledges financial support by Funda\c c\~ao Norte Riograndense de Pesquisa e Cultura (FUNPEC). This material is based upon work supported by the National Science Foundation under grant no. PHY-1654219 and OAC-1531814 and by the U.S. Department of Energy, Office of Science, Office of Nuclear Physics, within the framework of the Beam Energy Scan Theory (BEST) Topical Collaboration. The authors gratefully acknowledge the use of the Maxwell Cluster and the advanced support from the Center of Advanced Computing and Data Systems at the University of Houston.
\end{acknowledgments}

\appendix
\section*{Appendix}

In this appendix we give the details of the work presented in the main text and also provide some additional discussion about the topics covered. This is done in three main sections. In Section \ref{sec:supplEMD} we discuss in detail the holographic model we used, how the equations of motion are solved, and also how its parameters are fixed. We show the comparison to lattice thermodynamic data at zero chemical potential, extend the analysis to nonzero chemical potential, and discuss how to estimate the uncertainties in the location of the critical point in our holographic model. In Section \ref{sec:recon} we give additional details about the comparison of our model calculations at finite chemical potential to the available lattice calculations. In Section \ref{sec:rad} an analysis of the radius of convergence of the Taylor series for the thermodynamic quantities in powers of $\mu_B/T$ is performed.

\section{Holographic black hole engineering}
\label{sec:supplEMD}

The holographic gauge/gravity correspondence \cite{Maldacena:1997re,Gubser:1998bc,Witten:1998qj,Witten:1998zw} has been widely applied to obtain insight into the non-perturbative  behavior of different strongly correlated systems including the theory of strong interactions --- see \cite{CasalderreySolana:2011us,Adams:2012th} for broad reviews ---, condensed matter systems \cite{Donos:2012js,Hartnoll:2014lpa,Sachdev:2011wg}, and also quantum entanglement \cite{Nishioka:2009un,Maldacena:2013xja}.

Arguably, the most striking and general prediction made by holography is the small value obtained for the ratio between the shear viscosity and the entropy density of a strongly coupled quantum fluid described by a gravity dual in the bulk containing at most two derivatives in the gravity action (assuming spatial isotropy and translation invariance). Under such conditions, the gauge/gravity correspondence asserts that $\eta/s=1/4\pi$ \cite{Policastro:2001yc,Buchel:2003tz,Kovtun:2004de}, which is at least one order of magnitude smaller than perturbative QCD calculations for this ratio \cite{Arnold:2000dr,Arnold:2003zc}, being remarkably close to recent estimates obtained from comparisons between state-of-the-art hydrodynamic simulations of the quark-gluon plasma and heavy-ion data \cite{Bernhard:2016tnd}. Such a small value for $\eta/s$ is the defining property of the QGP produced by colliding heavy nuclei at RHIC and LHC \cite{Gyulassy:2004zy,Arsene:2004fa,Adcox:2004mh,Back:2004je,Adams:2005dq,Aad:2013xma,Heinz:2013th,Shuryak:2014zxa}.

The fact that $\eta/s$ in the strongly coupled regime of the QGP appears to be in the ballpark of the holographic result \cite{Kovtun:2004de} greatly increased the interest in applications of holographic models to the study of real time phenomena in the strongly coupled QGP, which are otherwise inaccessible to weak coupling techniques and are also very challenging to first principle lattice QCD simulations \cite{Meyer:2011gj} both at zero and nonzero baryon density \cite{Aarts:2011ax,Philipsen:2012nu}. On the other hand, most of the holographic studies conducted in this regard \cite{CasalderreySolana:2011us,Adams:2012th} have focused on studying properties of the so-called $\mathcal{N}=4$ super Yang-Mills (SYM) plasma, which turns out to be fairly different than the real-world QGP, especially in the crossover region \cite{Aoki:2006we,Borsanyi:2016ksw} where the QGP is highly nonconformal (see, for instance, the discussion in \cite{Rougemont:2016etk}).

More recently, bottom-up dilatonic gauge/gravity duals have been engineered with the aim of providing a realistic description of the physics of the nonconformal QGP \cite{Gubser:2008ny}. These constructions are mainly based on the coupling between the bulk metric field $g_{\mu\nu}$ and a real scalar field $\phi$ (which may be thought of as the dilaton), with the latter being responsible to break the conformal symmetry of the theory in the infrared regime, emulating the effects of a dynamically generated $\Lambda_\textrm{QCD}$ scale. This dynamical breaking of the conformal symmetry is controlled in the holographic model by the potential of the dilaton field, $V(\phi)$, which is a free function of the bottom-up construction that may be dynamically fixed by solving the Einstein-dilaton equations of motion with the constraint that the holographic equation of state at finite temperature ($T$) and zero baryon chemical potential ($\mu_B$) matches the corresponding lattice QCD result. Such a construction may then be employed to make predictions for a variety of observables relevant to characterize the physics of the QGP at zero baryon density \cite{Gubser:2008yx,Gubser:2008sz,Noronha:2009ud,Ficnar:2011yj,Ficnar:2012yu,Finazzo:2014cna,Janik:2015waa,Janik:2015iry,Attems:2016ugt,Attems:2016tby}. 

Effects due to a nonzero baryon chemical potential (or any other kind of Abelian chemical potential, such as the ones associated with the conservation of electric charge and strangeness) may be taken into account by adding a Maxwell field $A_\mu$ to the Einstein-dilaton action in the bulk, defining an Einstein-Maxwell-dilaton (EMD) model \cite{DeWolfe:2010he}. In this case, another free function is added to the model, corresponding to the coupling between the Maxwell and dilaton fields, $f(\phi)$. This coupling may be fixed by matching the holographically determined second order baryon susceptibility to the corresponding lattice result calculated at $\mu_B=0$. In this way, following the work of Ref.\ \cite{DeWolfe:2010he}, the EMD model becomes completely specified and may be used to provide predictions for many equilibrium and non-equilibrium observables at finite baryon density \cite{DeWolfe:2011ts,Rougemont:2015wca,Rougemont:2015ona,Finazzo:2015xwa,Knaute:2017opk,Rougemont:2017tlu}. More recently, an anisotropic version of the EMD model at finite magnetic field ($B$) and zero chemical potential has been developed and applied to determine the behavior of many physical quantities for the QGP across the $(T,B)$ plane \cite{Rougemont:2015oea,Finazzo:2016mhm,Critelli:2016cvq}.

Among the previous successes of bottom-up EMD holography applied to the QGP phenomenology, we highlight the following:

\begin{enumerate}[i.]

\item The EMD holographic model of Ref. \cite{Rougemont:2015wca} was shown in Refs. \cite{Finazzo:2013efa,Finazzo:2015xwa,Rougemont:2017tlu} to produce the results for the electric conductivity of the QGP which, among the results from different model calculations available in the literature (see e.g. the comparisons in Fig. 6 of \cite{Greif:2014oia} and in Fig. 4 of \cite{Greif:2016skc}), are the closest (both qualitatively and quantitatively) to the lattice QCD results with 2+1 flavors obtained in \cite{Aarts:2014nba}. Indeed, as discussed in Ref. \cite{Rougemont:2017tlu}, there is room for further improvements in the agreement between the EMD predictions for the electric conductivity and electric charge diffusion and the corresponding lattice QCD results from \cite{Aarts:2014nba}, once the latter are refined by taking the continuum limit and by also considering physical quark masses (as in the case of the lattice inputs used to fix the free parameters of the EMD holographic model).

\item In Ref. \cite{Rougemont:2017tlu} it was shown that the bulk viscosity of the EMD holographic model of Ref. \cite{Rougemont:2015wca} is very close, both qualitatively and quantitatively, to the result recently obtained in \cite{Bernhard:2017vql} through a Bayesian analysis of hydrodynamic simulations of the spacetime evolution of the QGP simultaneously matching different heavy ion experimental data.

\item The anisotropic EMD holographic model at finite temperature and magnetic field of Ref. \cite{Finazzo:2016mhm} was shown to quantitatively describe the anisotropic magnetized QCD equation of state and the magnetic field dependence of the pseudocritical crossover temperature obtained in state-of-the-art lattice QCD simulations at nonzero magnetic fields in \cite{Bali:2014kia}.

\item In Ref. \cite{Critelli:2016cvq}, the same anisotropic EMD model of Ref. \cite{Finazzo:2016mhm} was shown to quantitatively describe the renormalized Polyakov loop at finite magnetic field and the heavy quark entropy obtained in lattice QCD simulations in \cite{Bruckmann:2013oba,Endrodi:2015oba,Bazavov:2016uvm} for the QGP regime of the QCD phase diagram (i.e., for temperatures above the hadron gas regime).

\end{enumerate}

Moreover, as shown before in Figs. \ref{fig1} and \ref{recon_EOS} of the main text, the EMD model at finite temperature and baryon chemical potential constructed in the present work is able to quantitatively match state-of-the-art lattice results for the QCD equation of state with 2+1 flavors with physical quark masses up to the highest values of baryon chemical potential currently reached in lattice simulations \cite{Bazavov:2017dus}.\footnote{Note that the lattice simulations of Ref. \cite{Bazavov:2017dus} reach baryon chemical potentials up to $\mu_B\sim 600$ MeV.} The holographic equation of state at finite baryon density is not a result of any fitting procedure to lattice QCD data (which is only conducted at $\mu_B=0$ to fix the free parameters of the model, as aforementioned), but instead, it corresponds to a true prediction of the EMD model. Therefore, the quantitative agreement found in this work between the holographic equation of state and first principle lattice QCD results at finite baryon density corresponds to a highly nontrivial test of the phenomenological applicability of the EMD model to describe QCD data far from the region of the phase diagram where the free parameters of the bottom-up EMD model were fixed.

On the other hand, as it is well known, one cannot describe asymptotic freedom (setting in at very high energies in QCD) using gravity duals, since such constructions typically display strongly coupled instead of trivial ultraviolet fixed points. However, if there is a CEP in the QCD phase diagram at finite temperature and baryon density, as widely believed, it must be in the strongly coupled regime of QCD, otherwise it would has already been found in perturbative QCD calculations. Moreover, there are different model calculations which obtain a reduction in the shear viscosity times temperature to enthalpy density ratio as one increases the baryon density of the medium (see e.g., \cite{Denicol:2013nua} and also Refs. \cite{Rougemont:2015wca,Rougemont:2017tlu}), indicating that the QGP becomes more strongly coupled and closer to the perfect fluidity regime when it is doped with a nonzero baryon chemical potential. Consequently, the lack of asymptotic freedom in gravity duals is of no practical relevance for the phenomenological plausibility of the prediction we gave in the present work for the QCD CEP location. Instead, the quantitative agreement found between the holographic and lattice QCD equations of state at finite baryon density gives us confidence in the phenomenological reliability of such prediction.

The general form of the EMD action including finite $\mu_B$ effects employed in the present work, which we shall define in what follows, was first discussed in \cite{DeWolfe:2010he}. In that reference, now outdated lattice results for the equation of state and baryon susceptibility \cite{Karsch:2007dp} were used in the determination of the functions $V(\phi)$ and $f(\phi)$, which must then be revised to accommodate more precise lattice results. In \cite{Rougemont:2015wca} a new version of the EMD model was constructed which, contrary to the one originally devised in \cite{DeWolfe:2010he}, does not introduce any additional free parameters in the holographic model besides the ones already featured in the EMD action, making it a self-consistent gravitational setup. Furthermore, this new model employed more recent lattice QCD results for the equation of state \cite{Borsanyi:2012cr} and the dimensionless second order baryon susceptibility ($\chi_2$) \cite{Borsanyi:2011sw} with 2+1 flavors with physical quark masses. The new version of the EMD model parameters proposed in the present work (to be discussed in details in what follows) provides a much more precise description of state-of-the-art lattice results for $\chi_2$ and $s/T^3$ at $\mu_B=0$, where we match to the latest lattice QCD calculations from \cite{Borsanyi:2013bia}.\footnote{Note also that the results for the QCD equation of state at $\mu_B=0$ obtained by the HotQCD Collaboration in \cite{Bazavov:2014pvz} have now finally converged to the results of the Wuppertal-Budapest Collaboration \cite{Borsanyi:2013bia}.}

\subsection{EMD action and equations of motion}
\label{sec:EMDeoms}

The bulk EMD action is given by,
\begin{align}
S=\int_{\mathcal{M}_5} d^5x\,\mathcal{L} &= \frac{1}{2\kappa_5^2}\int_{\mathcal{M}_5} d^5x\,\sqrt{-g}\left[R-\frac{(\partial_\mu\phi)^2}{2}\right.\nonumber\\
&\left.-V(\phi) -\frac{f(\phi)F_{\mu\nu}^2}{4}\right],
\label{eq:EMDaction}
\end{align}
where $\kappa_5^2\equiv 8\pi G_5$ is the Newton's constant in five spacetime dimensions. The bulk action \eqref{eq:EMDaction} is complemented by some boundary terms which are, however, not necessary for the calculations done in the present work. In a bottom-up approach to the EMD model, one takes the dilaton potential $V(\phi)$ and the Maxwell-dilaton coupling $f(\phi)$ as free functions and there are also two free parameters, corresponding to the gravitational constant $\kappa_5^2$ and a characteristic energy scale, which we denote by $\Lambda$, used to convert physical observables calculated on the gravity side of the holographic duality in terms of inverse powers of the AdS radius $L$ to physical units (expressed in powers of MeV). By setting $L=1$ for simplicity, and introducing the energy scale $\Lambda$, we are simply exchanging the freedom to fix $L$ by the freedom to fix $\Lambda$ and, thus, the number of free parameters of the model is not augmented. In \ref{sec:EMDfits} we show how to fix these free parameters by matching lattice QCD results at $\mu_B=0$.

According to the holographic dictionary at finite temperature, thermal states of the 4-dimensional gauge theory with finite chemical potential are associated with charged black holes in the 5-dimensional bulk spacetime. We are interested here in static charged black hole backgrounds that are spatially isotropic and translationally invariant, which can be described by the following Ansatz for the EMD fields \cite{DeWolfe:2010he},
\begin{align}
ds^2&=e^{2A(r)} \left[ -h(r)dt^2+d\vec{x}^2 \right] +\frac{e^{2B(r)}dr^2}{h(r)},\nonumber\\
\phi&=\phi(r), \quad A=A_\mu dx^\mu=\Phi(r)dt, \label{eq:EMDansatz}
\end{align}
with the radial location of the black hole horizon given by the largest root of $h(r_H)=0$. We employ coordinates where the boundary of the asymptotically AdS$_5$ spacetime is at $r\rightarrow\infty$.

The equations of motion obtained by extremizing the action \eqref{eq:EMDaction} with respect to the EMD fields in the form given by the Ansatz \eqref{eq:EMDansatz} are given by \cite{DeWolfe:2010he},
\begin{align}
&\phi''(r)+\left[\frac{h'(r)}{h(r)}+4A'(r)-B'(r)\right]\phi'(r)-\frac{e^{2B(r)}}{h(r)}\left[ \frac{\partial V(\phi)}{\partial\phi}\right.\nonumber\\
&\left.-\frac{e^{-2[A(r)+B(r)]}\Phi'(r)^2}{2}\frac{\partial f(\phi)}{\partial\phi} \right]=0,\label{2.4}\\
&\Phi''(r)+\left[2A'(r)-B'(r)+\frac{d\left[\ln\left(f(\phi)\right)\right]}{d\phi}\phi'(r)\right]\Phi'(r)=0,\label{2.6}\\
&A''(r)-A'(r)B'(r)+\frac{\phi'(r)^2}{6}=0,\label{2.15}\\
&h''(r)+[4A'(r)-B'(r)]h'(r)-e^{-2A(r)}f(\phi)\Phi'(r)^2=0,\label{2.17}\\
&h(r)[24A'(r)^2-\phi'(r)^2]+6A'(r)h'(r)+2e^{2B(r)}V(\phi)\nonumber\\
&+e^{-2A(r)}f(\phi)\Phi'(r)^2=0,\label{2.16}
\end{align}
where the last equation is a useful constraint obtained by combining the independent components of Einstein's equations. Also, from the equations above, it is clear that the background function $B(r)$ has no dynamics. Indeed, due to reparametrization invariance of the radial coordinate, one has the freedom to fix $B(r)$ in order to simplify numerical calculations, as we are going to do in the next section. We also remark that there are two conserved charges in the radial direction, both associated with the EMD equations of motions, the Gauss charge $Q_G$, and the Noether charge $Q_N$ \cite{DeWolfe:2010he},
\begin{align}
Q_G(r)&=f(\phi)e^{2A(r)-B(r)}\Phi'(r),\nonumber\\
Q_N(r)&=e^{2A(r)-B(r)}[e^{2A(r)}h'(r)-f(\phi)\Phi(r)\Phi'(r)].
\label{2.18}
\end{align}
The equation of motion \eqref{2.6} for the gauge field $\Phi(r)$ may be written as $dQ_G/dr=0$, while the equation of motion \eqref{2.17} for the blackening function $h(r)$ may be written as $dQ_N/dr=0$.

\subsection{Numerical aspects and calculation of thermodynamic quantities}
\label{sec:EMDnum}

In order to numerically solve the EMD equations of motion and calculate physical observables we use two different sets of coordinates, both of them defined in the gauge where $B(r)=0$. We call coordinates with a tilde the ``standard coordinates'', while coordinates denoted without a tilde will be called ``numerical coordinates''. In the standard coordinates the blackening function goes to unity at the boundary, as usual, and one may calculate physical quantities such as the temperature or the entropy density using standard holographic formulas. On the other hand, for numerically solving the EMD equations of motion one needs to rescale these standard coordinates to specify definite values for some of the Taylor coefficients obtained by expanding the EMD fields near the black hole horizon, which is necessary to initialize the numerical integration of the equations of motion close to the horizon evolving them up to boundary of the asymptotically AdS$_5$ spacetime. This type of rescaling defines the numerical coordinates, as explained below.

\subsubsection{Thermodynamical functions in the standard coordinates}

Let us first review the derivation of the holographic formulas for the temperature ($T$), baryon chemical potential ($\mu_B$), entropy density ($s$), and baryon charge density ($\rho_B$) in the standard coordinates (denoted with a tilde). As mentioned above, we work in the $\tilde{B}(\tilde{r})=0$ gauge, in terms of which the EMD fields \eqref{eq:EMDansatz} take the form
\begin{align}
d\tilde{s}^2&=e^{2\tilde{A}(\tilde{r})}\left[-\tilde{h}(\tilde{r})d\tilde{t}^2+d\vec{\tilde{x}}^2\right]+ \frac{d\tilde{r}^2}{\tilde{h}(\tilde{r})},\nonumber\\
\tilde{\phi}&=\tilde{\phi}(\tilde{r}),\quad\tilde{A}=\tilde{A}_\mu d\tilde{x}^\mu=\tilde{\Phi}(\tilde{r})d\tilde{t}.
\label{2.19}
\end{align}
Physical quantities in the gauge theory are usually obtained from the far-from-the-horizon, near-boundary behavior of the bulk fields. One may obtain the ultraviolet behavior of these fields by first considering $\tilde{\phi}(\tilde{r}\to\infty)\to 0$, $V(0)=-12$, $f(0)=\textrm{const}$, $\tilde{h}(\tilde{r}\to\infty)\to 1$, and then substituting these results into the EMD equations of motion, solving them close to the boundary $\tilde{r}\to\infty$ in terms of $\tilde{A}(\tilde{r})$ (with the requirement that the background metric goes back to the AdS$_5$ geometry at the boundary) and $\tilde{\Phi}(\tilde{r})$. After this is done, one may consider the backreaction of these fields into the dynamics of the dilaton field, as one slowly starts to go into the interior of the bulk, by plugging these results back into the EMD equations of motion and solving them for $\tilde{\phi}(\tilde{r})$ with the dilaton potential now truncated at quadratic order. This backreacted process may be repeated to obtain the following ultraviolet expansion of the EMD fields close to the boundary in the standard coordinates, first derived in \cite{DeWolfe:2010he},
\begin{align}
\tilde{A}(\tilde{r})&=\tilde{r}+\mathcal{O}\left(e^{-2\nu\tilde{r}}\right),\nonumber \\
\tilde{h}(\tilde{r})&=1+\mathcal{O}\left(e^{-4\tilde{r}}\right),\nonumber \\
\tilde{\phi}(\tilde{r})&=e^{-\nu\tilde{r}}+\mathcal{O}\left(e^{-2\nu\tilde{r}}\right),\nonumber \\
\tilde{\Phi}(\tilde{r})&=\tilde{\Phi}_0^{\textrm{far}}+\tilde{\Phi}_2^{\textrm{far}}e^{-2\tilde{r}}+ \mathcal{O}\left(e^{-(2+\nu)\tilde{r}}\right),
\label{2.20}
\end{align}
where $\nu\equiv d-\Delta$, $d=4$ being the number of spacetime dimensions of the dual gauge theory. $\Delta=(d+\sqrt{d^2+4m^2})/2$ is the scaling dimension of the gauge theory operator dual to the bulk dilaton field and $m$ is the mass of the dilaton obtained by Taylor expanding the dilaton potential close to the boundary. For the potential we shall consider here (to be discussed in section \ref{sec:EMDfits}) $\Delta < d$ and, thus, the dilaton is dual to a relevant gauge theory operator responsible for triggering a renormalization group flow from an ultraviolet fixed point towards a nonconformal state as one goes to the infrared regime of the quantum gauge theory.

Now we are ready to obtain standard holographic formulas for the thermodynamical variables. The temperature in the gauge theory equals the Hawking's temperature of the black hole,
\begin{align}
T=\frac{\sqrt{-g'_{\tilde{t}\tilde{t}} g^{\tilde{r}\tilde{r}}\,'}}{4\pi}\biggr|_{\tilde{r}=\tilde{r}_H}\!\!\!\!\!\!\Lambda= \frac{e^{\tilde{A}(\tilde{r}_H)}}{4\pi}|\tilde{h}'(\tilde{r}_H)|\Lambda,
\label{2.21}
\end{align}
where we have introduced the energy scale $\Lambda$ (to be fixed in section \ref{sec:EMDfits}) to express $T$ in physical units (correspondingly, any gauge/gravity observable with energy dimension $p$ will be multiplied by $\Lambda^p$ when expressed in physical units). Note that such procedure, contrary to the one employed in \cite{DeWolfe:2010he}, naturally respects the fact that dimensionless combinations of dimensionful observables should be independent of the units used to measure them; this is clearly violated when one introduces different energy scales to express different dimensionful observables in powers of MeV as done in \cite{DeWolfe:2010he}, besides also artificially augmenting the number of free parameters of the holographic model. The entropy density in the gauge theory is holographically associated with the area of the bulk black hole horizon by means of the well-known Bekenstein-Hawking formula \cite{Bekenstein:1973ur,Hawking:1974sw},
\begin{align}
s=\frac{A_H}{4G_5V}\Lambda^3=\frac{2\pi}{\kappa_5^2}e^{3\tilde{A}(\tilde{r}_H)}\Lambda^3.
\label{2.22}
\end{align}
By following the holographic dictionary, one extracts the baryon chemical potential in the gauge theory from the boundary value of the bulk gauge field,
\begin{align}
\mu_B=\lim_{\tilde{r}\rightarrow\infty}\tilde{\Phi}(\tilde{r})\Lambda=\tilde{\Phi}_0^{\textrm{far}}\Lambda,
\label{2.23}
\end{align}
while the baryon charge density is obtained from the boundary value of the radial momentum conjugate to the bulk Maxwell field,
\begin{align}
\rho_B=\lim_{\tilde{r}\rightarrow\infty} \frac{\partial\mathcal{L}}{\partial\left(\partial_{\tilde{r}}\tilde{\Phi}\right)} \Lambda^3
=\frac{Q_G(\tilde{r}\rightarrow\infty)}{2\kappa_5^2}\Lambda^3
=-\frac{\tilde{\Phi}_2^{\textrm{far}}}{\kappa_5^2}\Lambda^3.
\label{2.24}
\end{align}

\subsubsection{Thermodynamical functions in the numerical coordinates}

In order to numerically solve the EMD equations of motion, we now shift to numerical coordinates defined by the following procedure. We first consider near horizon Taylor expansions of the bulk EMD fields, $X(r)=\sum_{n=0}^\infty X_n(r-r_H)^n$, where $X=\left\{A,h,\phi,\Phi\right\}$. Then, by rescaling the holographic coordinate one may fix $r_H=0$; $h_0=0$ follows from the fact that the blackening function has a simple zero at the black hole horizon; $h_1=1$ may be fixed by rescaling the time coordinate while $A_0=0$ may be fixed by rescaling the spacetime coordinates parallel to the boundary, $(t,\vec{x})$, by a common factor. Moreover, since $dt$ has infinite norm at the horizon, if $\Phi(r_H)=\Phi_0\neq 0$ one would obtain an ill defined Maxwell field at the black hole horizon, which imposes $\Phi_0=0$ for consistency. With the near horizon Taylor coefficients $h_0$, $h_1$, $A_0$, and $\Phi_0$ determined as above, one may find the remaining Taylor expansion coefficients as functions of two initial conditions, $(\phi_0,\Phi_1)$, by solving the EMD equations of motion order by order in the aforementioned expansions.

One avoids the singular point of the differential equations at the horizon, $r_H=0$, by starting the numerical integration at a slightly shifted position, for instance, at $r_{\textrm{start}}=10^{-8}$. Additionally, second order near-horizon Taylor expansions may be employed, $X(r_{\textrm{start}})=X_0+X_1r_{\textrm{start}}+X_2r_{\textrm{start}}^2+\mathcal{O}(r_{\textrm{start}}^3)$, to numerically integrate the EMD equations of motion from the shifted horizon $r_{\textrm{start}}$ up to the boundary, which may be numerically parametrized by some ultraviolet cutoff, e.g., $r_{\textrm{max}}=2$, corresponding to a value of the radial coordinate where the numerically generated black hole backgrounds have already reached the ultraviolet fixed point corresponding to the AdS$_5$ spacetime. The six unknown second order Taylor coefficients, $h_2$, $A_1$, $A_2$, $\phi_1$, $\phi_2$, and $\Phi_2$ may be then determined as functions of the initial conditions $(\phi_0,\Phi_1)$ by substituting the second order near horizon expansions into the differential equations \eqref{2.4} --- \eqref{2.16} and setting to zero each power of $r_{\textrm{start}}$ in the resulting algebraic system. The near horizon boundary conditions necessary to initialize the numerical integration of the EMD equations of motion \eqref{2.4} --- \eqref{2.17} are then given by $X(r_{\textrm{start}})$ and $X'(r_{\textrm{start}})$.

We remark that for each possible value of the initial condition $\phi_0$ there is a bound on the maximum value allowed for the initial condition $\Phi_1$ above which the numerical solutions fail to be asymptotically AdS$_5$. This bound may be derived by noting that in the $B(r)=0$ gauge the equation of motion \eqref{2.15} gives $A''(r)=-\phi'(r)^2/6\le 0$, implying that $A(r)$ is a concave function of the holographic coordinate. As done in \cite{DeWolfe:2010he,Rougemont:2015wca}, we restrict our calculations in the present work to positive values of the initial condition $\phi_0$, which is enough to generate a holographic phase diagram in close agreement to what is uncovered in state-of-the-art lattice QCD simulations. Taking also into account that for asymptotically AdS$_5$ geometries the background function $A(r)$ must increase for large values of $r$, it turns out that $A(r)$ must be a monotonically increasing function. This implies that the derivative of $A(r)$ at the horizon must be positive, $A_1>0$. By plugging the near horizon expansions into the constraint \eqref{2.16} and evaluating it at the black hole horizon one obtains,
\begin{align}
A_1=-\frac{1}{6}\left[2V(\phi_0)+f(\phi_0)\Phi_1^2\right].
\label{2.26}
\end{align}
We work with a negative-definite dilaton potential $V(\phi)$ and a positive-definite Maxwell-dilaton coupling $f(\phi)$ and, since for asymptotically AdS$_5$ spacetimes one must have $A_1>0$, Eq.\ \eqref{2.26} leads to the following bound \cite{DeWolfe:2010he},
\begin{align}
\Phi_1<\sqrt{-\frac{2V(\phi_0)}{f(\phi_0)}}\equiv\Phi_1^{\textrm{max}}(\phi_0).
\label{2.27}
\end{align}

As mentioned before, physical quantities on the gauge theory side of the correspondence are usually calculated from the near boundary, far from the horizon behavior of the bulk fields. In the numerical coordinates, the ultraviolet behavior of these fields reads \cite{DeWolfe:2010he},
\begin{align}
A(r)&=\alpha(r)+\mathcal{O}\left(e^{-2\nu\alpha(r)}\right),\nonumber\\
h(r)&=h_0^{\textrm{far}}+\mathcal{O}\left(e^{-4\alpha(r)}\right),\nonumber\\
\phi(r)&=\phi_A e^{-\nu\alpha(r)}+\mathcal{O}\left(e^{-2\nu\alpha(r)}\right),\nonumber\\
\Phi(r)&=\Phi_0^{\textrm{far}}+\Phi_2^{\textrm{far}}e^{-2\alpha(r)}+ \mathcal{O}\left(e^{-(2+\nu)\alpha(r)}\right),\label{2.28}
\end{align}
where $\alpha(r)= A_{-1}^{\textrm{far}}r+A_0^{\textrm{far}}$. Evaluation of the constraint \eqref{2.16} at the boundary gives $A_{-1}^{\textrm{far}}=1/\sqrt{h_0^{\textrm{far}}}$. By equating the radially conserved Gauss charge in Eq.\ \eqref{2.18} evaluated at the horizon and at the boundary, one finds
\begin{align}
\Phi_2^{\textrm{far}}=-\frac{\sqrt{h_0^{\textrm{far}}}}{2f(0)}f(\phi_0)\Phi_1.
\label{2.30}
\end{align}
For the calculations carried out here, one just needs to obtain the behavior of a few ultraviolet expansion coefficients of the EMD fields close the boundary. These coefficients are $h_0^{\textrm{far}}$, $\Phi_0^{\textrm{far}}$, $\Phi_2^{\textrm{far}}$, and $\phi_A$. One may reliably fix $h_0^{\textrm{far}}=h(r_{\textrm{max}})$ and $\Phi_0^{\textrm{far}}=\Phi(r_{\textrm{max}})$, since the blackening function and the Maxwell field quickly reach the values corresponding to a conformal theory. With $h_0^{\textrm{far}}$ now determined, $\Phi_2^{\textrm{far}}$ may be obtained from Eq.\ \eqref{2.30}. The ultraviolet coefficient $\phi_A$ is more complicated to fix in a reliable way because it multiplies an exponentially decreasing function. In the present work, we employ the same procedure originally devised in \cite{Finazzo:2016mhm}, which is more general and efficient than the one used in \cite{Rougemont:2015wca}. Both procedures give the same results for the dilaton potential and Maxwell-dilaton coupling used in \cite{Rougemont:2015wca}; however, for the dilaton potential and Maxwell-dilaton coupling used in the present work (to be discussed in section \ref{sec:EMDfits}), the procedure used in \cite{Rougemont:2015wca} can only reliably cover a very narrow region of the plane of initial conditions $(\phi_0,\Phi_1)$, while the numerical procedure used in \cite{Finazzo:2016mhm} to obtain $\phi_A$ provides a reliable covering of a much wider region. The reliability in the extraction of $\phi_A$ is checked by comparing the numerical results for the dilaton field close to the boundary with its analytical near boundary expansion given in Eq.\ \eqref{2.28}. We use the ultraviolet fitting profile $\phi_{\textrm{fit}}^{\textrm{UV}}(r)=\phi_A e^{-\nu\alpha(r)}$, defined within the adaptive interval $r\in[r_{\textrm{IR}}(\phi_0,\Phi_1)=\phi^{-1}(10^{-3}),r_{\textrm{UV}}(\phi_0,\Phi_1)=\phi^{-1}(10^{-5})]$, to fit the numerically generated dilaton field $\phi(r)$ close the boundary, with the ultraviolet coefficient $\phi_A$ emerging as the outcome of this fitting procedure.

Finally, in order to directly evaluate the thermodynamical functions in Eqs.\ \eqref{2.21} --- \eqref{2.24} in terms of the numerically generated black hole backgrounds, one needs to relate the standard and the numerical coordinates of the $B(r)=0$ gauge. This may be done by setting $\tilde{\phi}(\tilde{r})=\phi(r)$, $d\tilde{s}^2=ds^2$, $\tilde{\Phi}(\tilde{r})d\tilde{t}=\Phi(r)dt$ and by comparing the ultraviolet asymptotics given in Eqs.\ \eqref{2.20} and \eqref{2.28}, from which it follows that \cite{DeWolfe:2010he},
\begin{align}
&\tilde{r}=\frac{r}{\sqrt{h_0^{\textrm{far}}}}+A_0^{\textrm{far}}-\ln(\phi_A^{1/\nu}),\quad
\tilde{A}(\tilde{r})=A(r)-\ln(\phi_A^{1/\nu}),\\
&\vec{\tilde{x}}=\phi_A^{1/\nu}\vec{x}, \quad
\tilde{t}=\phi_A^{1/\nu}\sqrt{h_0^{\textrm{far}}}\,t,\quad
\tilde{h}(\tilde{r})=\frac{h(r)}{h_0^{\textrm{far}}},\\
&\tilde{\Phi}(\tilde{r})\!=\!\frac{\Phi(r)}{\phi_A^{1/\nu}\sqrt{h_0^{\textrm{far}}}},\,\,
\tilde{\Phi}_0^{\textrm{far}}\!=\!\frac{\Phi_0^{\textrm{far}}}{\phi_A^{1/\nu}\sqrt{h_0^{\textrm{far}}}},\,\, \tilde{\Phi}_2^{\textrm{far}}\!=\!\frac{\Phi_2^{\textrm{far}}}{\phi_A^{3/\nu}\sqrt{h_0^{\textrm{far}}}}.
\label{2.37}
\end{align}
With this one finally obtains
\begin{align}
T&=\frac{1}{4\pi\phi_A^{1/\nu}\sqrt{h_0^{\textrm{far}}}}\,\Lambda,\label{2.40}\\
\mu_B&=\frac{\Phi_0^{\textrm{far}}}{\phi_A^{1/\nu}\sqrt{h_0^{\textrm{far}}}}\,\Lambda,\label{2.41}\\
s&=\frac{2\pi}{\kappa_5^2\,\phi_A^{3/\nu}}\,\Lambda^3,\label{2.42}\\
\rho_B&=-\frac{\Phi_2^{\textrm{far}}}{\kappa_5^2\,\phi_A^{3/\nu}\sqrt{h_0^{\textrm{far}}}}\,\Lambda^3.
\label{2.43}
\end{align}

\subsection{Fixing the free parameters of the EMD model via black hole engineering}
\label{sec:EMDfits}

In order to dynamically fix the free parameters of the bottom-up EMD model, we match the holographic entropy density and the second order baryon susceptibility to the corresponding lattice QCD results with $2+1$ flavors and physical quark masses calculated at $\mu_B=0$. We already have in Eqs.\ \eqref{2.40} --- \eqref{2.43} what is needed to deal with the equation of state. Regarding the dimensionless baryon susceptibility $\chi_2$, one may derive a simple integral expression for it at vanishing baryon density (the details of this derivation may be found in \cite{DeWolfe:2010he,Rougemont:2015wca}),
\begin{align}
\chi_2(\mu_B=0)=\frac{1}{16\pi^2} \frac{s}{T^3} \frac{1}{f(0)\int_{r_H}^\infty dr\, e^{-2A(r)}f^{-1}(\phi(r))},
\label{2.52}
\end{align}
which is to be evaluated using the neutral black hole backgrounds defined at $\mu_B=0$ obtained by setting the initial condition $\Phi_1$ to zero. In numerical calculations, one replaces in Eq.\ \eqref{2.52} $r_H\mapsto r_{\textrm{start}}$ and $\infty\mapsto r_{\textrm{max}}$.

Each pair of initial conditions $(\phi_0,\Phi_1)$ generates a 5-dimensional black hole geometry that is asymptotically AdS$_5$ corresponding, through the holographic dictionary given by Eqs.\ \eqref{2.40} - \eqref{2.43}, to a thermodynamical state with definite values of $(T,\mu_B,s,\rho_B)$ in the strongly coupled gauge theory. Then, by spanning many different values of $(\phi_0,\Phi_1)$ one generates an ensemble of charged black hole backgrounds, each one of them corresponding to a point in the phase diagram of the holographic model.

The free parameters of the model are fixed at $\mu_B=0$ by lattice QCD inputs for the equation of state and second order baryon susceptibility such that the EMD results for these observables at vanishing baryon density are not to be taken as predictions of the model - they stem from a simultaneous dynamical fitting procedure used to constrain the free parameters of the bottom-up construction. In this context, we say that this procedure corresponds to \emph{holographic black hole engineering} \cite{Critelli:2016cvq}, i.e., black hole solutions are engineered to display the relevant properties of the QGP found on the lattice at $\mu_B=0$. On the other hand, everything calculated in the holographic model at nonzero $\mu_B$, as well as other physical quantities calculated at $\mu_B=0$ which were not used to fix the free parameters of the EMD setup such as transport coefficients, follow as bonafide predictions of our model.

\begin{figure*}
\begin{center}
\begin{tabular}{c}
\subfigure[]{\includegraphics[width=0.4\textwidth]{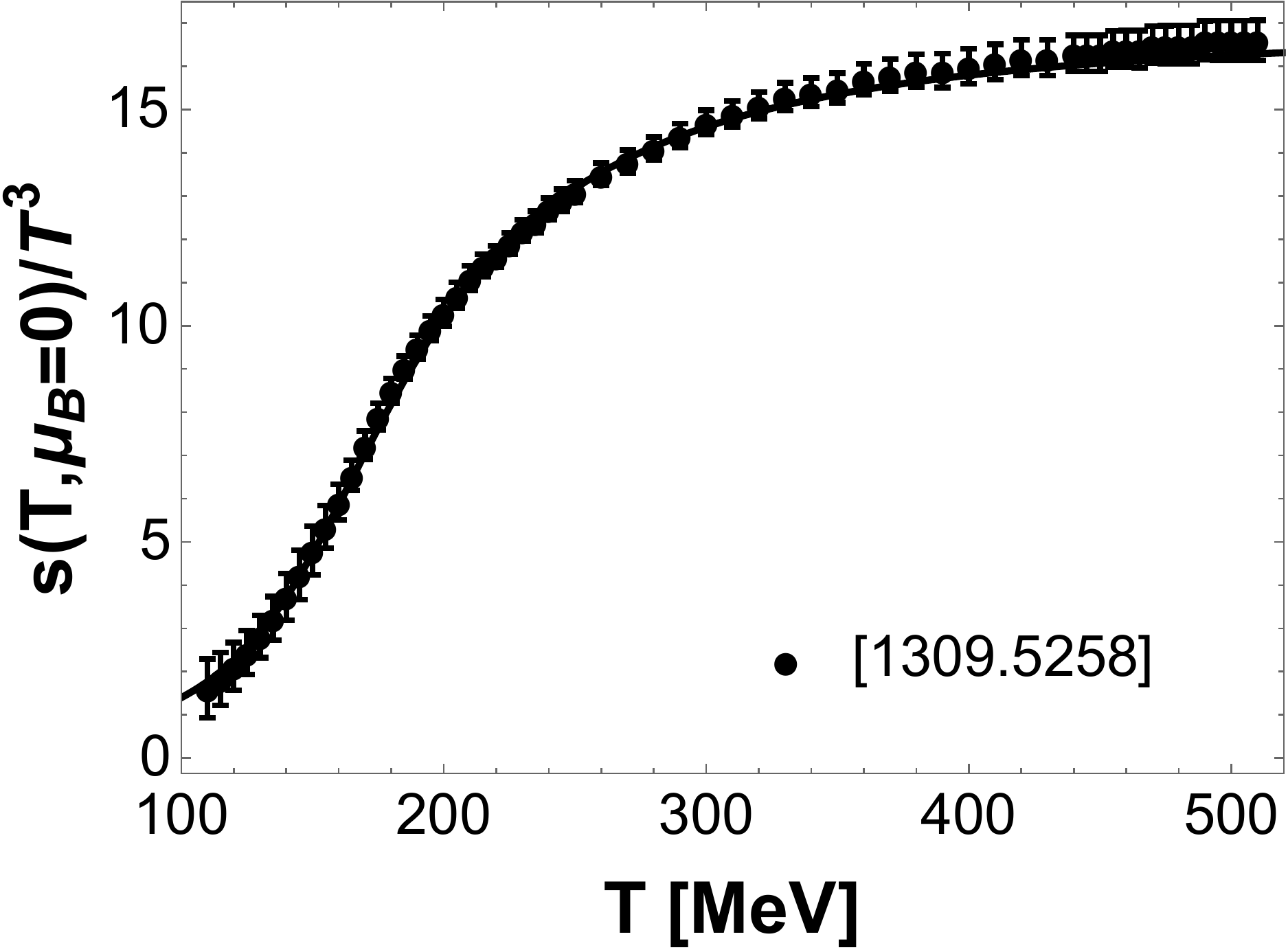}} 
\end{tabular}
\begin{tabular}{c}
\subfigure[]{\includegraphics[width=0.4\textwidth]{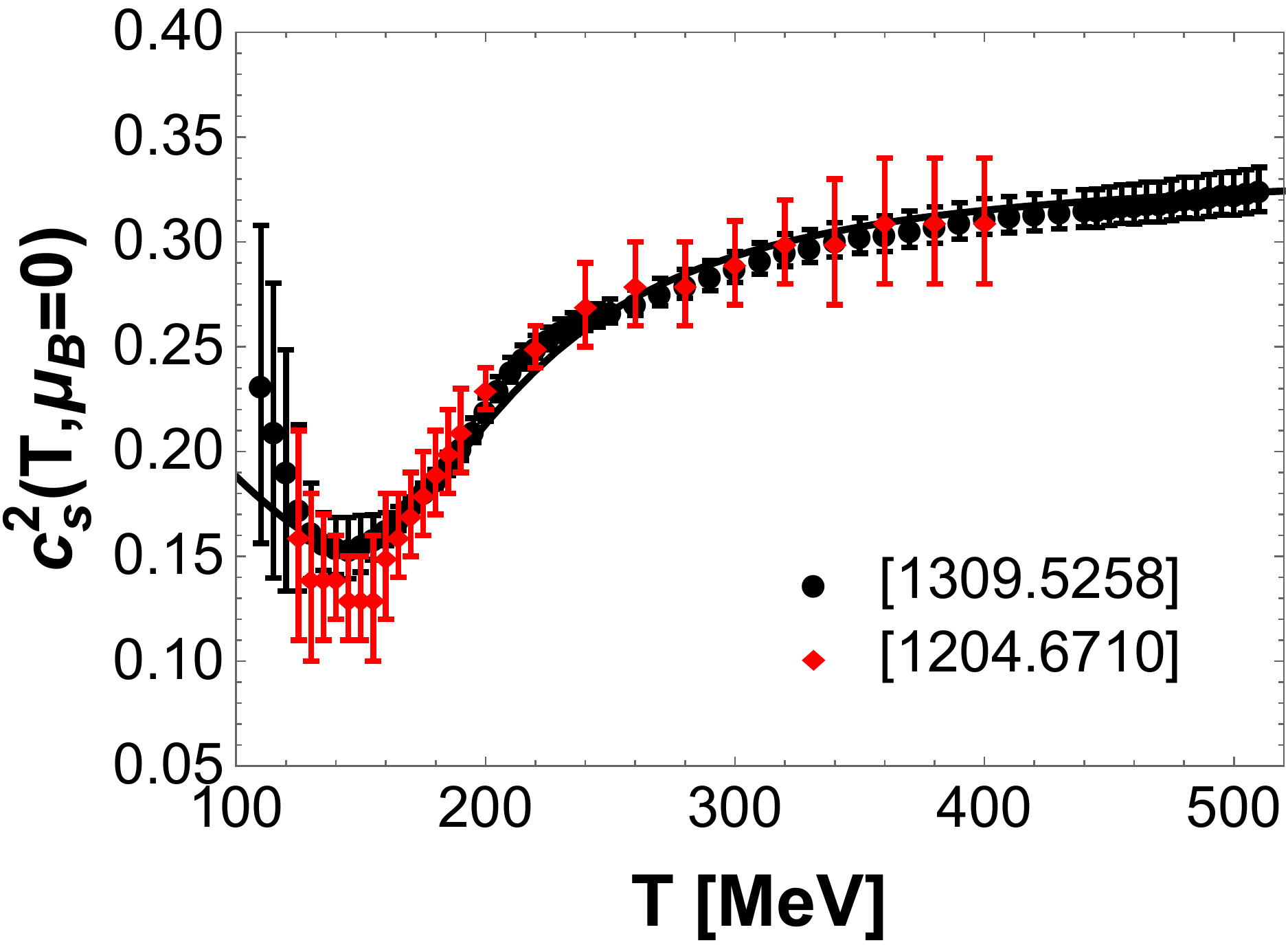}} 
\end{tabular}
\begin{tabular}{c}
\subfigure[]{\includegraphics[width=0.4\textwidth]{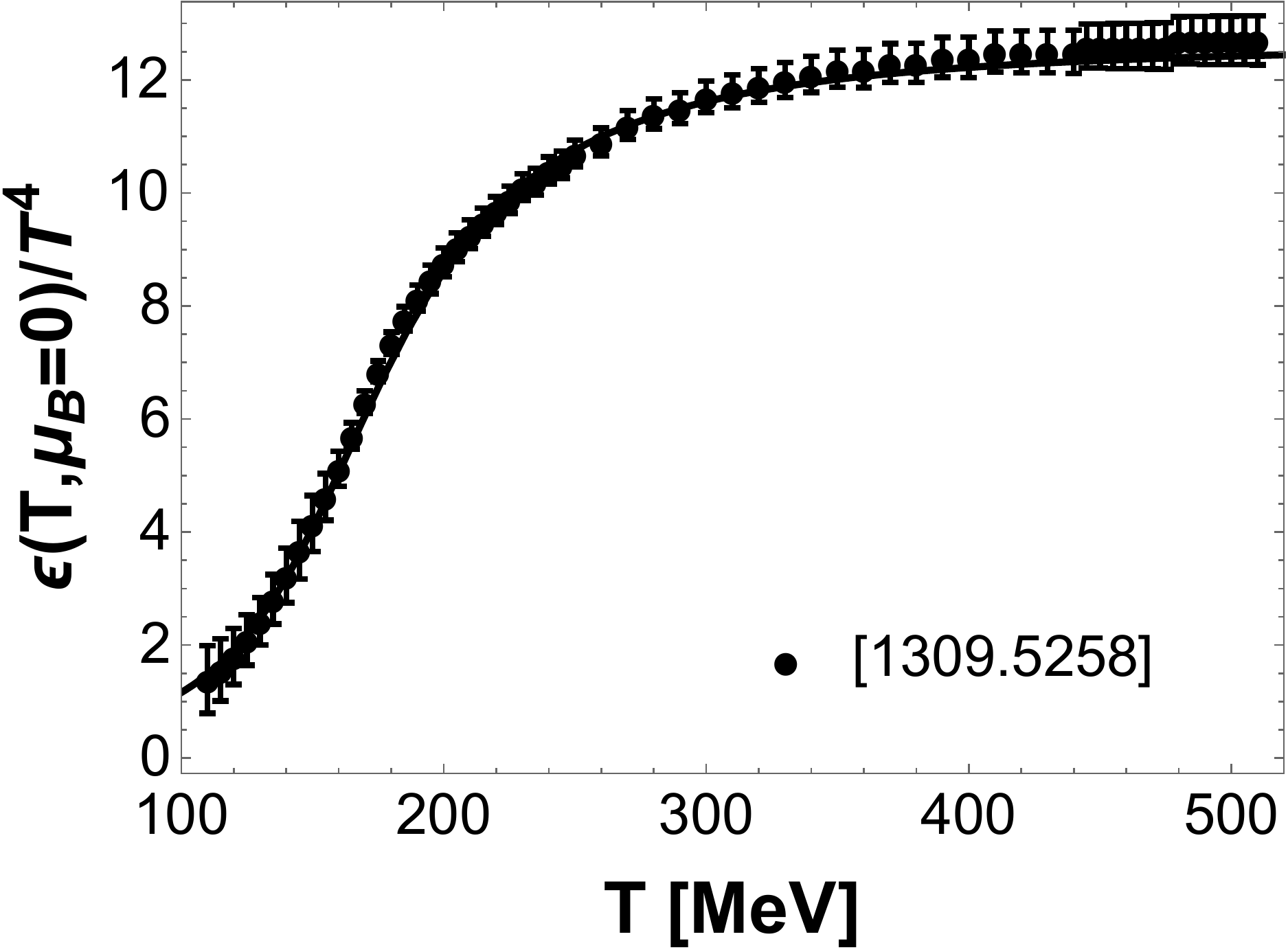}} 
\end{tabular}
\begin{tabular}{c}
\subfigure[]{\includegraphics[width=0.4\textwidth]{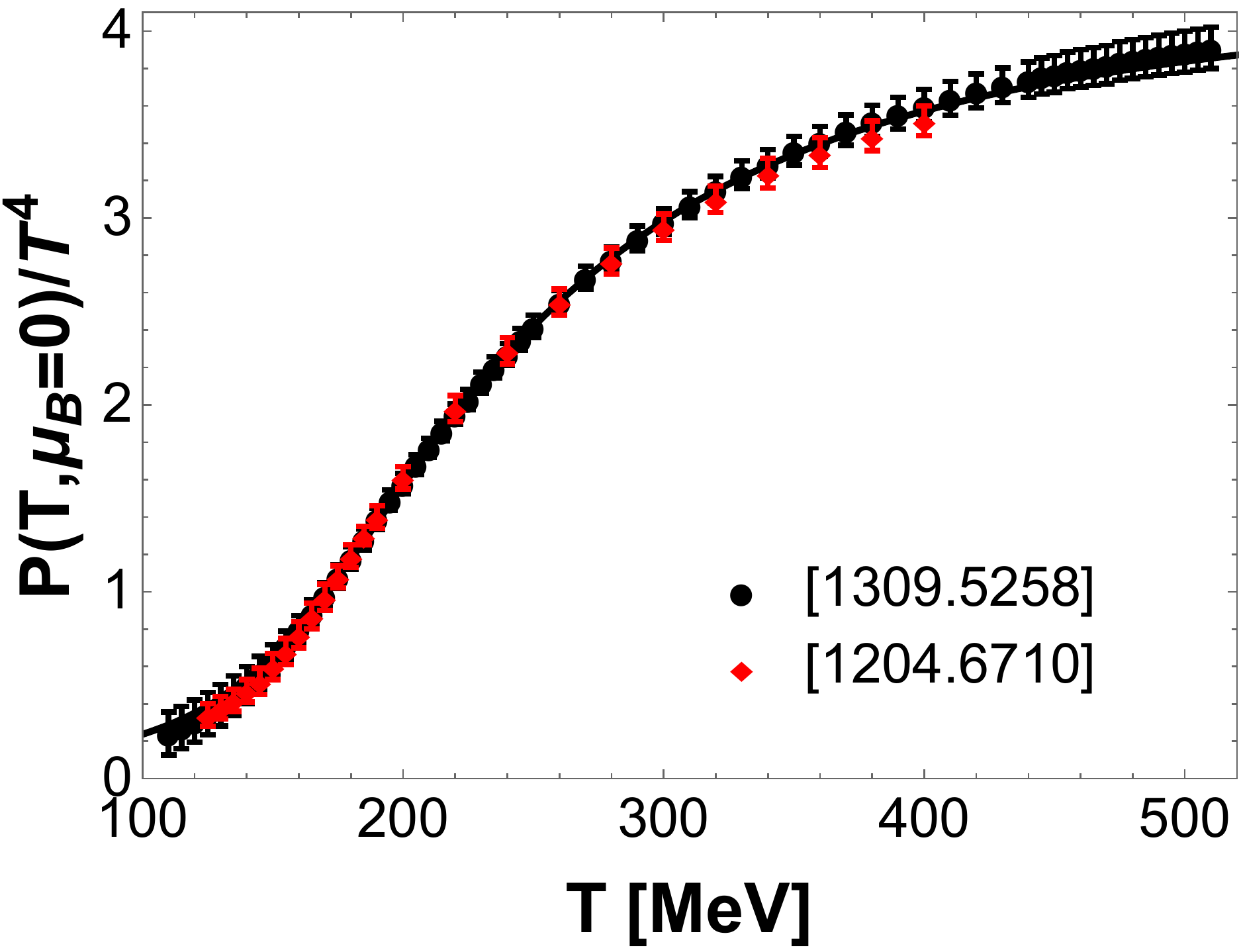}} 
\end{tabular}
\begin{tabular}{c}
\subfigure[]{\includegraphics[width=0.4\textwidth]{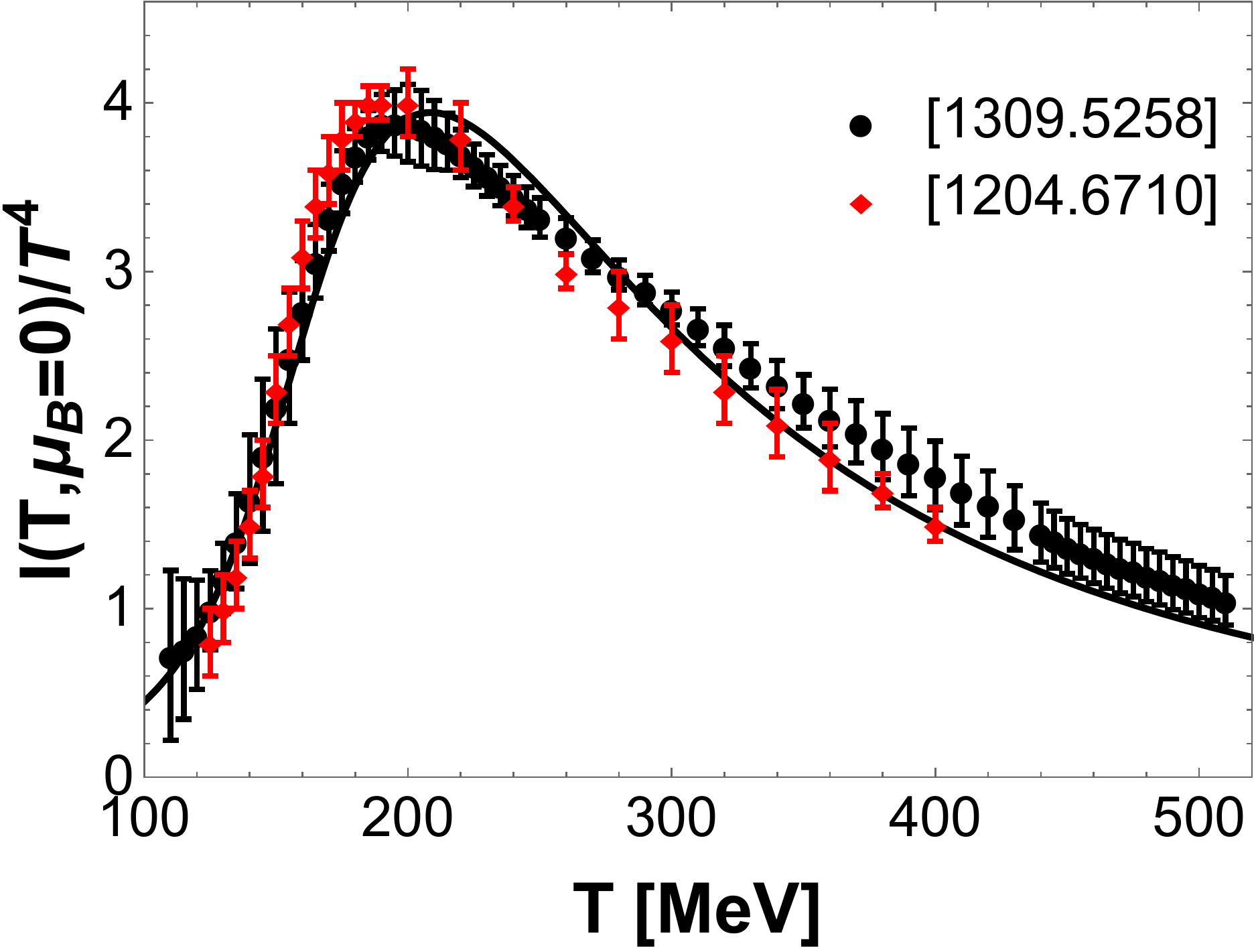}} 
\end{tabular}
\end{center}
\caption{(Color online) Thermodynamics  at $\mu_B=0$ from lattice QCD results \cite{Borsanyi:2013bia} compared to our holographic model (we also plot in red the older calculations from \cite{Borsanyi:2012cr} for a comparison with the latest results from \cite{Borsanyi:2013bia}): (a) entropy density, (b) speed of sound squared, (c) energy density $\epsilon$, (d) pressure $P$, (f) trace anomaly $I = \epsilon-3P$.}
\label{fig:EMDfits}
\end{figure*}

In this paper, we simultaneously match the holographic results for the entropy density ($s/T^3$) and second order baryon susceptibility ($\chi_2$) to state-of-the-art lattice QCD results for these quantities computed using $2+1$ flavors and physical quark masses from Refs.\ \cite{Borsanyi:2011sw,Borsanyi:2013bia,Bellwied:2015lba}. The other thermodynamic quantities follow directly using well known thermodynamic identities. The free parameters of the EMD holographic model fixed in this way are given by
\begin{align}
V(\phi)&=-12\cosh(0.63\,\phi)+0.65\,\phi^2-0.05\,\phi^4+0.003\,\phi^6,\nonumber\\
\kappa_5^2&=8\pi G_5=8\pi(0.46), \quad \Lambda=1058.83\,\textrm{MeV},\nonumber\\
f(\phi)&=\frac{\textrm{sech}\left(c_1 \phi + c_2\phi^2\right)}{1+c_3}+\frac{c_3}{1+c_3} \textrm{sech}(c_4\phi),
\label{eq:EMDfits}
\end{align}
where $c_1 = -0.27$, $c_2 = 0.4$, $c_3 = 1.7$, and $c_4=100$, with the corresponding results displayed in Fig.\ \ref{fig:EMDfits} (the excellent agreement obtained for $\chi_2$ was already shown in Fig.\ \ref{fig1} of the main text). We note that $V(\phi)$, $\kappa_5^2$, and $\Lambda$ were originally fixed in \cite{Finazzo:2016mhm}. We also remark that the effective mass of the dilaton field obtained from $V(\phi)$, $m^2\approx -3.46$, satisfies the Breitenlohner-Freedman bound for massive scalar fields defined on asymptotically AdS$_5$ spacetimes \cite{Breitenlohner:1982jf,Breitenlohner:1982bm}. The scaling dimension of the gauge theory operator dual to the dilaton field is $\Delta\approx 2.73$, which corresponds to a relevant deformation as anticipated in previous sections. 

We close this section by remarking that the holographic pressure was calculated here by integrating the entropy density with respect to the temperature by using the following approximation (this is actually a pressure difference),
\begin{align}
P(T,\mu_B=0)\approx\int_{T_{\textrm{low}}}^T d\bar{T}\, s(\bar{T},\mu_B=0),
\label{eq:pressuredif}
\end{align}
where we took $T_{\textrm{low}}=70$ MeV. Clearly, this approximation will no longer be adequate to determine the pressure when $T \to T_{\textrm{low}}$. However, for the values of $T$ we used to present the EMD results for the pressure in this work, this approximation gives fairly stable results. We checked, for instance, that the results obtained using $T_{\textrm{low}}=10$ MeV are to a very good approximation the same obtained using $T_{\textrm{low}}=70$ MeV in \eqref{eq:pressuredif}. The reason why we employ $T_{\textrm{low}}=70$ MeV throughout the present work to calculate the pressure is because, for the grid of initial conditions we were able to numerically generate covering the region of the critical point of the EMD phase diagram (to be discussed in the next section), there are not too many points with $T < 70$ MeV. Points at lower values of $T$ may be generated by changing the borders of the rectangle of initial conditions in the $(\phi_0,\Phi_1)$ plane but, in this case, we were not able to adequately cover the region of the $(T,\mu_B)$ plane where the critical point of the model is located.

\subsection{Holographic thermodynamics at finite baryon density}
\label{sec:EMDthermo}

Using Eqs.\ \eqref{2.40} - \eqref{2.43} one is able to calculate several thermodynamical quantities at finite temperature and baryon density. The internal and free energy densities at finite $\mu_B$ are given by, respectively,
\begin{align}
\epsilon(s,\rho_B)&=Ts-P+\mu_B\rho_B,\\
\mathcal{F}(T,\mu_B)&=-P(T,\mu_B)=\epsilon(s,\rho_B)-Ts-\mu_B\rho_B.
\end{align}
From the above equations one obtains the following differential relations,
\begin{align}
d\epsilon(s,\rho_B)&=Tds+\mu_B d\rho_B,\\
d\mathcal{F}(T,\mu_B)&=-dP(T,\mu_B)=-sdT-\rho_B d\mu_B,
\end{align}
such that at fixed $\mu_B$,
\begin{align}
dP(T,\textrm{fixed}\,\mu_B)=sdT,
\end{align}
and the speed of sound squared at a fixed value of $\mu_B$ is given by
\begin{align}
c_s^2(T,\mu_B)&=\frac{dP}{d\epsilon}\biggr|_{\mu_B}\nonumber\\
&=\left(\frac{T}{s}\frac{\partial s(T,\mu_B)}{\partial T}\biggr|_{\mu_B} +\frac{\mu_B}{s}\frac{\partial \rho_B(T,\mu_B)}{\partial T}\biggr|_{\mu_B}\right)^{-1}.
\label{cs2mu}
\end{align}
This equation was used to obtain the transition line corresponding to the minimum of $c_s^2$ used in the main text. For completeness, we remind the reader that the expression for the trace anomaly at finite $\mu_B$ includes the effect of the baryon density 
\begin{align}
I(T,\mu_B)&=\epsilon(T,\mu_B)-3P(T,\mu_B)\nonumber\\
&=Ts(T,\mu_B)+\mu_B\rho_B(T,\mu_B)-4P(T,\mu_B).
\label{Imu}
\end{align}

\begin{figure*}
\begin{center}
\begin{tabular}{cc}
\includegraphics[width=0.45\textwidth]{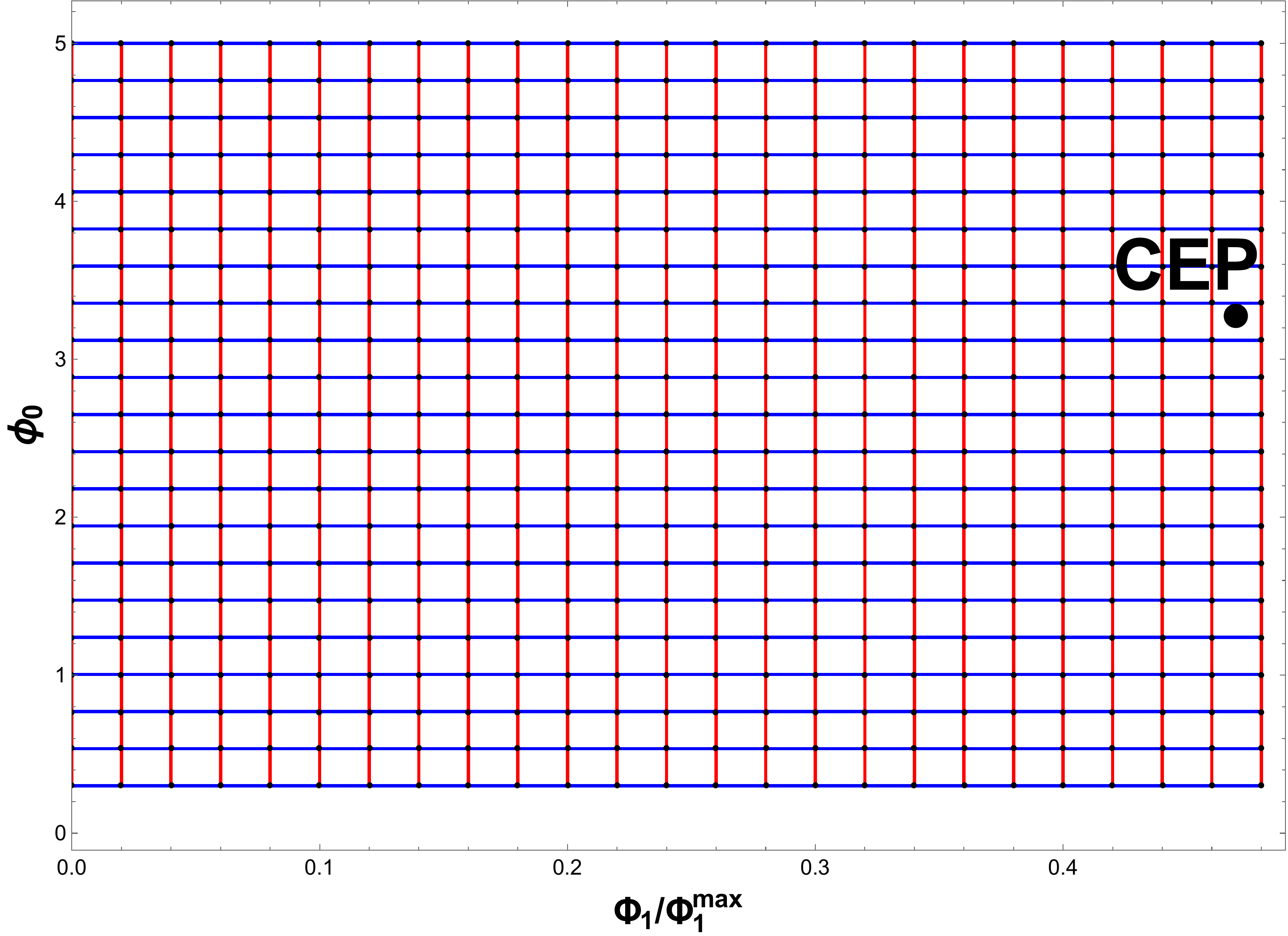} &
		\includegraphics[width=0.45\textwidth]{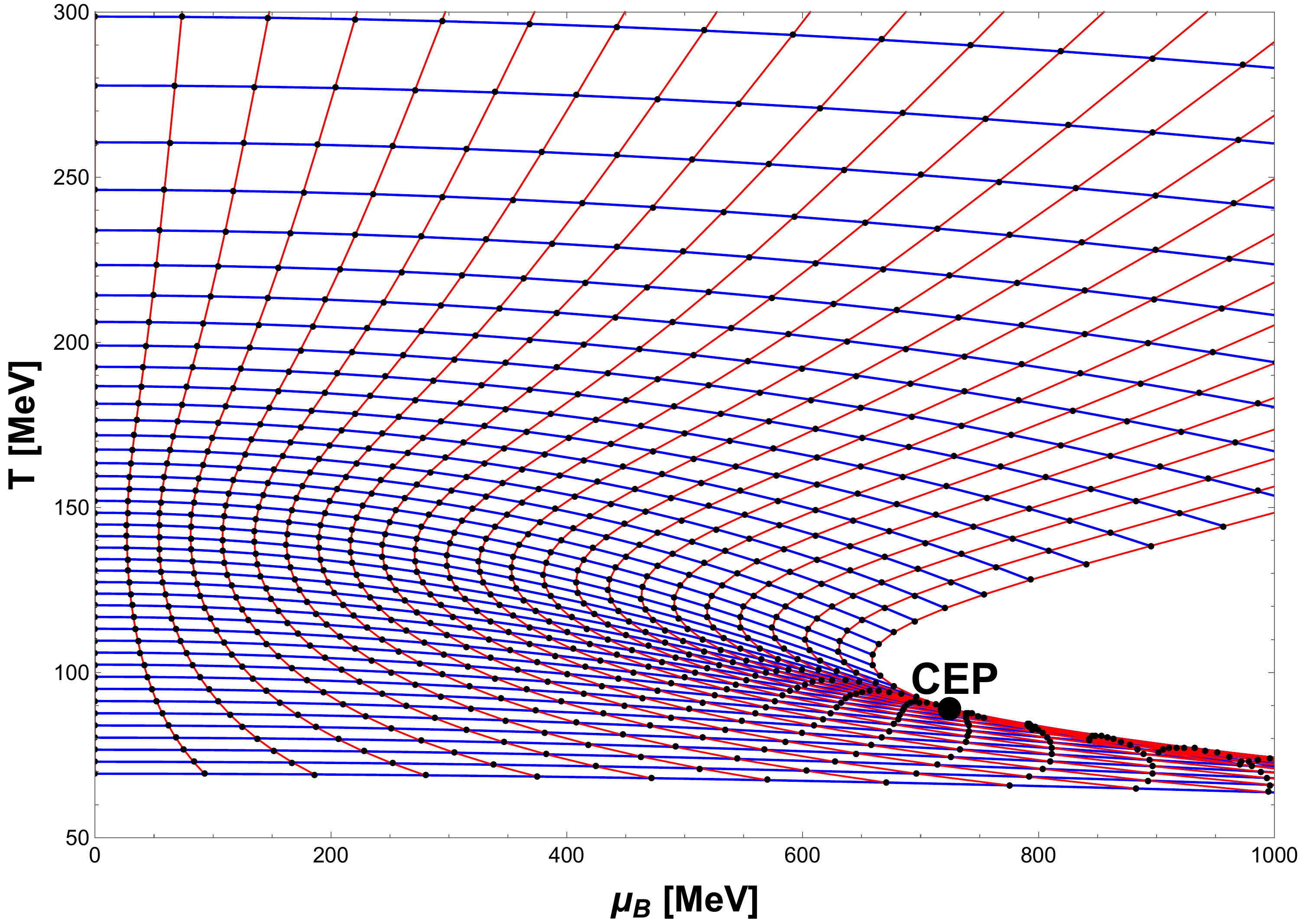}
		\end{tabular}
\end{center}
\caption{(Color online) The plot shows how an equally spaced $\left(\phi_0,\Phi_1\right)$ grid is mapped into an irregular grid in the $(T,\mu_B)$ plane generated by the black hole solutions used in this work. The critical point is depicted in both plots.}
\label{fig:EMDgridCEP}
\end{figure*}

For the results presented in this paper, we numerically generated an ensemble containing altogether $2 \times 10^6$ charged black holes with initial conditions spanning the rectangle defined by $\phi_0\in[0.3,5]$ and $\Phi_1\in[0,0.48]\Phi_1^{\textrm{max}}(\phi_0)$. This rectangle of initial conditions is broad enough to reveal the location of the critical point of the present EMD setup and also to provide the necessary information for the calculation of the higher order baryon susceptibilities presented in the main text. Fig.\ \ref{fig:EMDgridCEP} shows how an equally spaced $\left(\phi_0,\Phi_1\right)$ grid is mapped into an irregular grid in the $(T,\mu_B)$ plane generated by the black hole solutions used in this work.

In Fig.\ \ref{recon_EOS} of the main text, we show a comparison between the holographic EMD predictions for the equation of state at finite baryon density and state-of-the-art lattice QCD results obtained for $\mu_B/T \simeq 2$ from Ref.\ \cite{Bazavov:2017dus}: there is very good agreement between both calculations, which gives us confidence that the present EMD model can provide the first reliable estimate for the location of the critical endpoint in the baryon rich regime of the QCD phase diagram.

In Fig.\ \ref{fig:EMDgridCEP} (right) we locate the CEP of the EMD model at $(T^{\textrm{CEP}},\,\mu_B^{\textrm{CEP}})=(89,724)$ MeV. This estimate was obtained by determining the location of the numerical divergence of the second order baryon susceptibility and checking that the baryon charge density acquires an infinite slope at the CEP, as expected. 

\begin{figure}
\begin{center}
\includegraphics[width=0.45\textwidth]{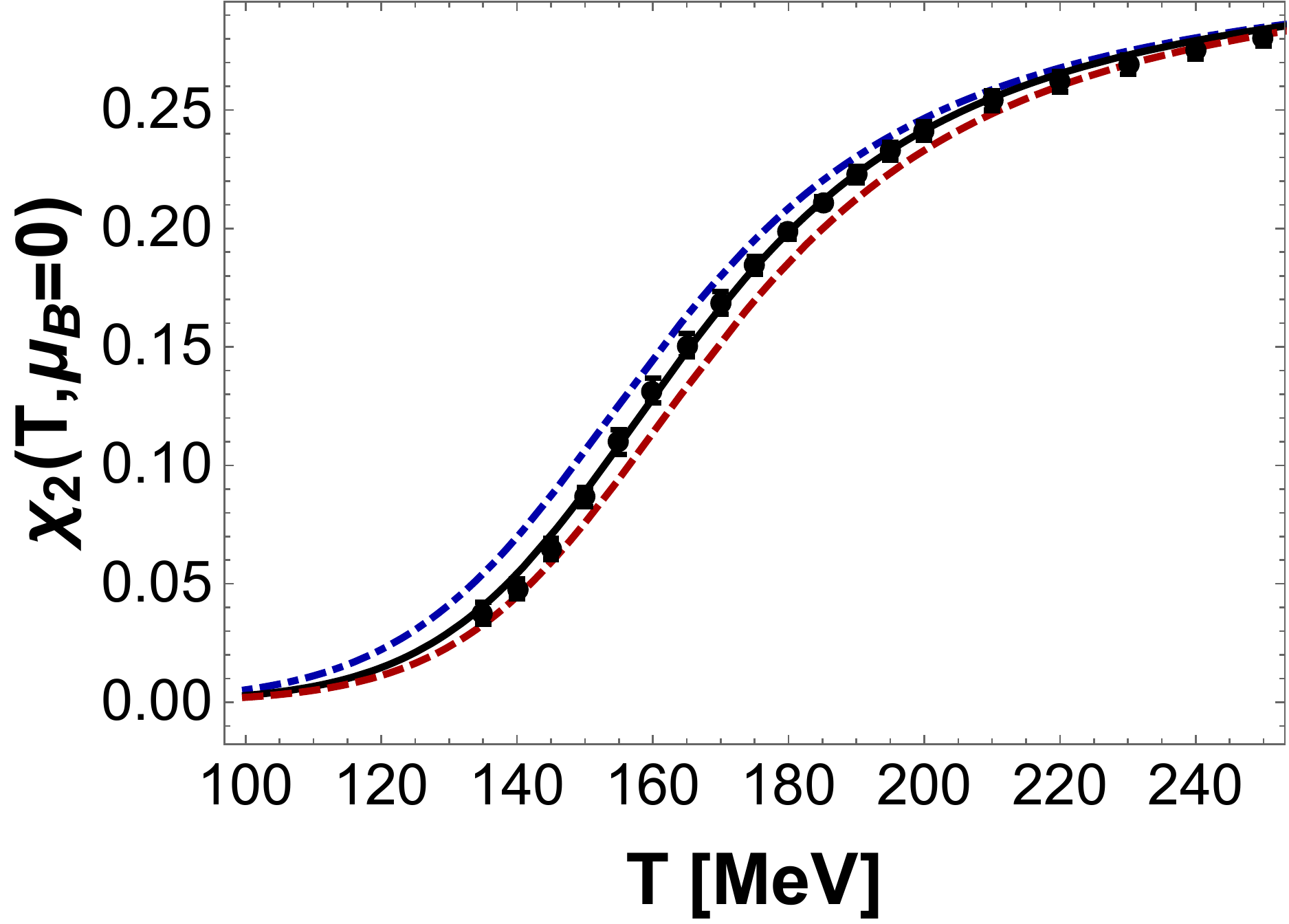}
\end{center}
\caption{(Color online) Examples that illustrate how variations of the model parameters, performed to assess the effects of the small error bars in the lattice calculations \cite{Borsanyi:2011sw}, change the holographic result for the second order baryon susceptibility. The dashed red and dot-dashed blue curves are generated by varying either $c_1$ and $c_2$ in \eqref{eq:EMDfits}. The solid black curve represents our best set of parameters used in this work that gives a CEP at $T_{CEP}=89$ MeV and $\mu_B^{CEP}=724$ MeV.}
\label{fig:errorbandchi2}
\end{figure}

In order to better understand how the location (and ultimately the presence) of the critical point depends on how the baryon density effects are encoded in the parameters of the model, we varied the parameters of $f(\phi)$ in such a way to estimate the effect of the (very small) error bars coming from the lattice calculation of $\chi_2$ \cite{Borsanyi:2011sw}. A particular example of this time consuming study is illustrated by the red and blue lines in Fig.\ \ref{fig:errorbandchi2}, shown in comparison to the lattice points and the solid black curve corresponding to our best set of parameters discussed around Eq.\ \eqref{eq:EMDfits}. The curve that is below the lattice results (dashed red line) was created setting $c_1= -0.189$ keeping the other parameters fixed in \eqref{eq:EMDfits}, while the upper curve (dot-dashed blue line) was obtained changing $c_2$ to $0.36$ keeping the other parameters fixed. This analysis, together with other many tests, has led to the estimate quoted in the main text that a variation of the parameters of the model can shift $T_{CEP}$ by at most $13\%$ and $\mu_B^{CEP}$ by at most $5\%$, if one requires that the $\chi_2$ computed holographically is still broadly consistent with the very small uncertainty in the lattice calculations.

We close this section by mentioning that our prediction for the QCD critical point in the phase diagram is located at a smaller $T$ and larger $\mu_B$ than other previous estimates using different approaches ranging from lattice QCD-based analyses \cite{Fodor:2004nz,Datta:2016ukp}, an experimentally-driven finite-size scaling analysis \cite{Lacey:2014wqa}, and Dyson-Schwinger models \cite{Xin:2014ela,Fischer:2014ata}. See the reviews \cite{Stephanov:2004wx} and \cite{Luo:2017faz} for other relevant references (in this regard, \cite{Fu:2015amv,Fu:2016tey,Fan:2016ovc} are examples of recent studies of QCD critical phenomena in effective models). Finally, we would like to remark that we did not consider non-equilibrium effects \cite{Berdnikov:1999ph} in our analysis of critical phenomena and their possible signatures in heavy ion collisions. This has been the subject of many interesting studies, e.g., \cite{Jiang:2015hri,Herold:2016uvv,Mukherjee:2016kyu}, and it certainly requires further investigation on the holographic side (for recent studies of near and far-from-equilibrium dynamics in a holographic top-down model at finite density with a critical point see \cite{Finazzo:2016psx,Critelli:2017euk}).

\section{Reconstruction of the QCD equation of state at finite $\mu_B$}
\label{sec:recon}

Due to how well the black hole engineering approach reproduces the higher order susceptibilities at $\mu_B=0$ calculated using lattice QCD (see Fig.\ \ref{recon_EOS} of the main text), it uniquely allows us to investigate different methods that use this information to reconstruct the QCD equation of state at finite $\mu_B$ as well as to find out how many susceptibilities $\chi_n$ are needed to accurately reconstruct the equation of state out to a certain value of $\mu_B/T$. While the order of the expansion was already discussed in the main text surrounding Fig.\ \ref{recon_EOS} for the pressure and $\rho_B$, it is also important to study different ways within which the series itself may be reconstructed. For instance, a Pad\'e approximant generates poles in the complex $\mu_B/T$ plane and it may have the advantage of showing early indications of a critical point. The Pad\'e reconstructions for $\rho_B(T,\mu_B)$ including terms up to $\mathcal{O}(\mu_B^3)$ and $\mathcal{O}(\mu_B^4)$ are given by
\begin{widetext}
\begin{align}
\rho_B(T,\mu_B)=\frac{\chi_2\left(\frac{\mu_B}{T}\right)+\frac{10(\chi_4)^2-3\chi_2\chi_6}{60\chi_4} \left(\frac{\mu_B}{T}\right)^3}{1-\chi_6\left(\frac{\mu_B}{T}\right)^2/(20\chi_4)},\quad
\rho_B(T,\mu_B)=\frac{\chi_2\left(\frac{\mu_B}{T}\right)+\frac{70(\chi_4)^3-42\chi_2\chi_4\chi_6+ 3(\chi_2)^2\chi_8}{42\left(10(\chi_4)^2-3\chi_2\chi_6\right)}\left(\frac{\mu_B}{T}\right)^3}
{1+\frac{-7\chi_4\chi_6+\chi_2\chi_8}{14\left(10(\chi_4)^2-3\chi_2\chi_6\right)}\left(\frac{\mu_B}{T}\right)^2+  \frac{21(\chi_6)^2-10\chi_4\chi_8}{840\left(10(\chi_4)^2-3\chi_2\chi_6\right)}\left(\frac{\mu_B}{T}\right)^4},
\label{eqn:pade}
\end{align}
\end{widetext}
respectively.

\begin{figure*}
	\begin{center}
		\includegraphics[width=0.8\textwidth]{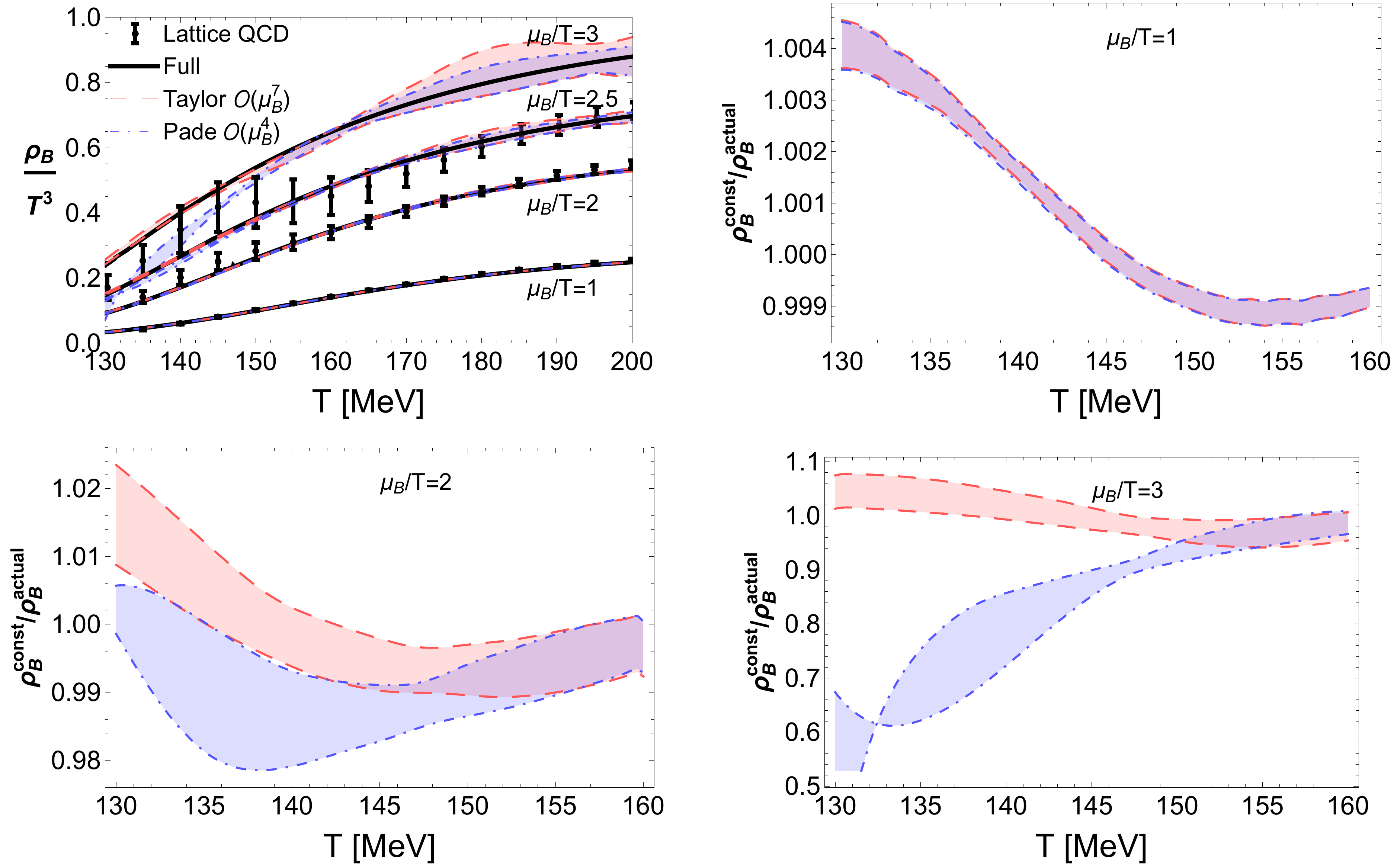}
	\end{center}
	\caption{\label{rat_recon}(Color online) The reconstructed baryon density via either a Taylor series or a Pad\'e approximation. In this figure the result including terms up to $\chi_8$ (top left) is compared to the full baryon density calculated in our holographic model and also on the lattice \cite{Bazavov:2017dus}. The other panels show the ratio of the reconstructed baryon density from the Taylor expansion and Pad\'e approximation calculated up to $\chi_8$ over $\rho_B$ calculated directly from the black hole model defined using three different values of $\mu_B/T=$ 1 (top right), 2 (bottom left), and 3 (bottom right).}
\end{figure*}

In Fig.\ \ref{rat_recon} (top left) a comparison between the directly calculated baryon density, $\rho_B$, and the reconstructed $\rho_B$ using either the usual Taylor series or \eqref{eqn:pade} are shown. At large $\mu_B$ the Taylor series converges more quickly to the actual $\rho_B$ and gives a reasonable approximation up to almost $\mu_B/T\sim 3$. Looking at the ratios of the reconstructed $\rho_B$ to the actual $\rho_B$ one can see that up to $\mu_B/T\sim 2$  both methods work reasonably well and the error is at most only $1-2\%$.  However, when $\mu_B/T\sim 3$ for the Taylor series there is less than a $10\%$ error while the Pad\'e approximation has a significant deviation at $\mu_B/T\sim 3$ with an error up to 40$\%$ in the low temperature region. In this case, the Taylor series is more adequate to reconstruct the equation of state and this will be used in the calculations below.

\begin{figure*}
	\begin{tabular}{cc}
		\includegraphics[width=0.5\textwidth]{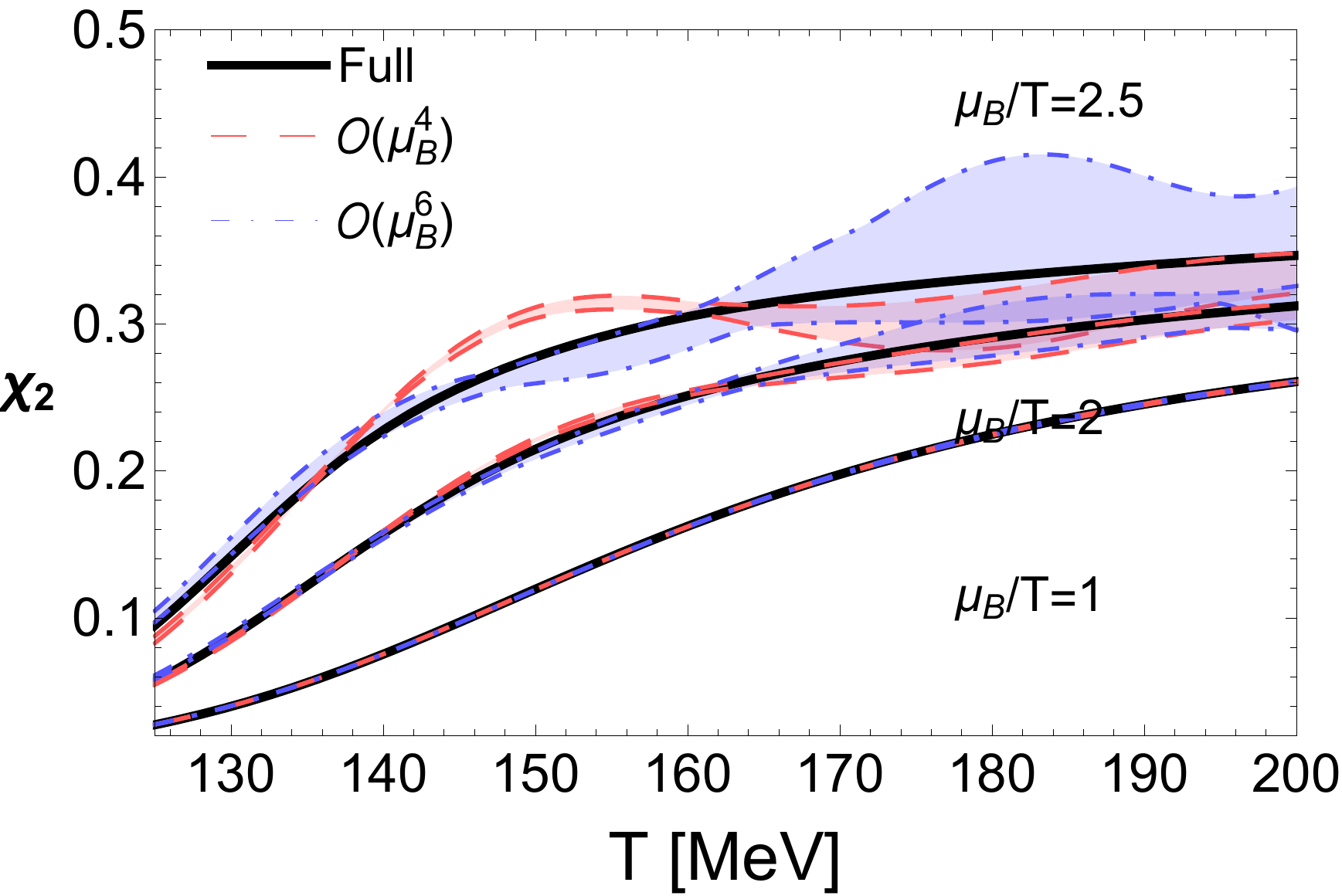} & 	
		\includegraphics[width=0.5\textwidth]{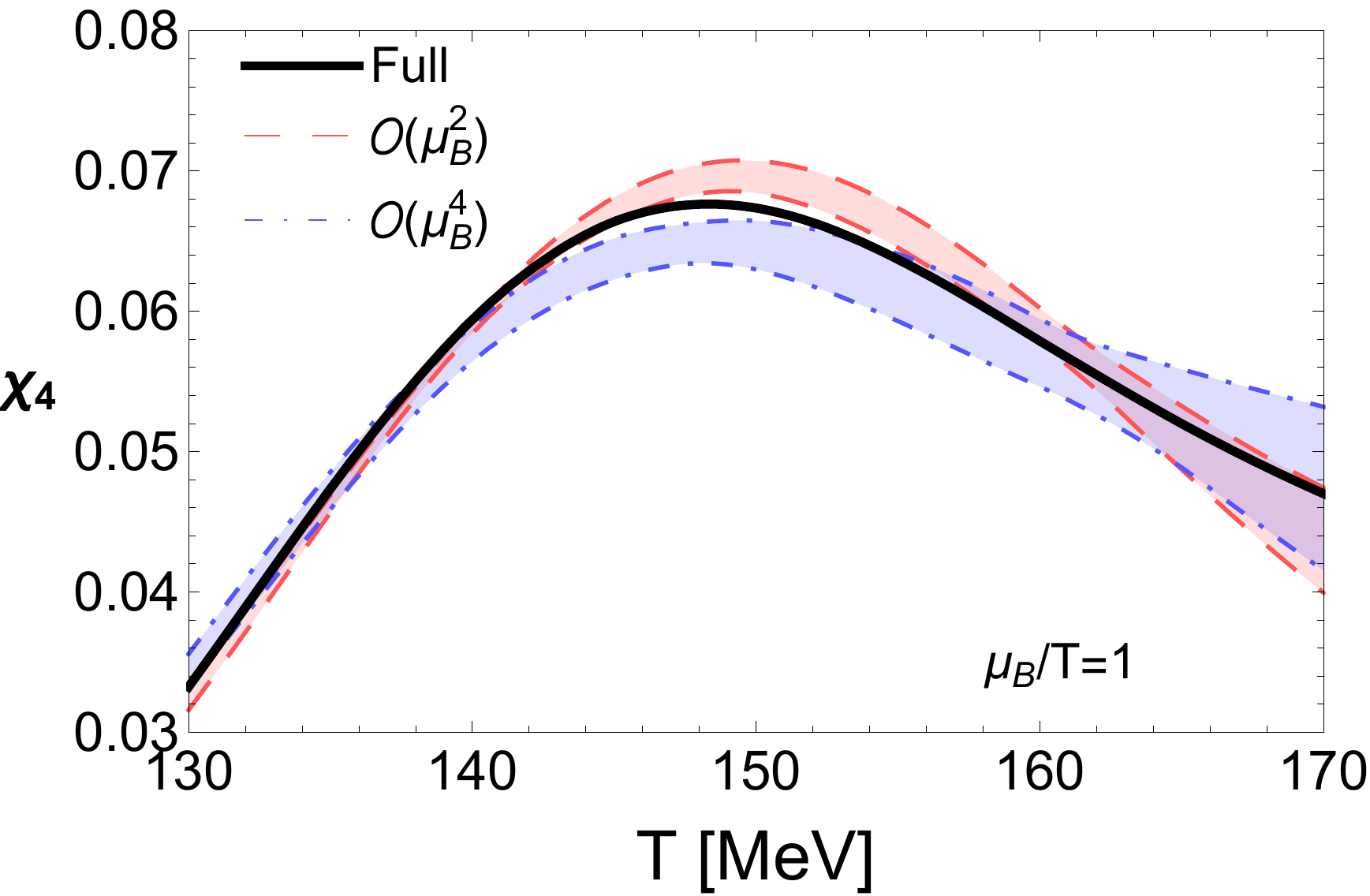}
	\end{tabular}
	\caption{\label{chi2_recon}(Color online) Comparison between the full holographic results for $\chi_2$ and $\chi_4$ and their reconstructed Taylor series: $\chi_2$ at $\mu_B/T=$ 1, 2, and 2.5 (left) and $\chi_4$ at $\mu_B/T=$ 1 (right) calculated directly from the black hole model compared to the Taylor expansion including terms up to $\mathcal{O}(\mu_B^4)$ and $\mathcal{O}(\mu_B^6)$ for $\chi_2$ and up to $\mathcal{O}(\mu_B^2)$ and $\mathcal{O}(\mu_B^4)$ for $\chi_4$.}
\end{figure*}

Next, the truncation order of the Taylor series needed to reconstruct $\chi_2(T,\mu_B)$ and $\chi_4(T,\mu_B)$ is studied, motivated by the fact that higher order susceptibilities are more strongly affected by the critical point ($\chi_2$ diverges at the critical point and, therefore, any potential peak displayed by $\chi_2$ is relevant for investigations about critical phenomena in QCD). In Fig.\ \ref{chi2_recon} (left) the reconstructed  $\chi_2$ is shown across different values of $\mu_B/T$ where there is a reasonable good description up to $\mu_B/T\sim 2$ using terms up to $\mathcal{O}(\mu_B^{6})$. The slope of $\mu_B/T\sim 2$ at $\mathcal{O}(\mu_B^{4})$ artificially stiffens due to the limited number of terms in the Taylor series, which can lead to misleading conclusions.  At larger $\mu_B/T$ the curvature is distorted even at $\mathcal{O}(\mu_B^{6})$.  Even though it is not surprising that the validity of the Taylor series for $\chi_2$ is limited to a smaller region of $\mu_B/T$ compared to $\rho_B$, this highlights the need to extend the current lattice calculations to even higher order susceptibilities. As a matter of fact, the series for $\chi_4$ has an even smaller range in $\mu_B/T$ and already struggles to reproduce the directly calculated  $\chi_4$ at $\mu_B/T=1$, as shown in Fig.\ \ref{chi2_recon} (right). Therefore, Fig.\ \ref{chi2_recon} shows that additional higher order susceptibilities at $\mu_B=0$ beyond $\chi_8$ are needed to simultaneously obtain reasonable descriptions of $\chi_2 (T,\mu_B)$ and $\chi_4 (T,\mu_B)$ using Taylor expansions for $\mu_B/T>1$.

\begin{figure}[ht!]
\centering
\includegraphics[width=0.45\textwidth]{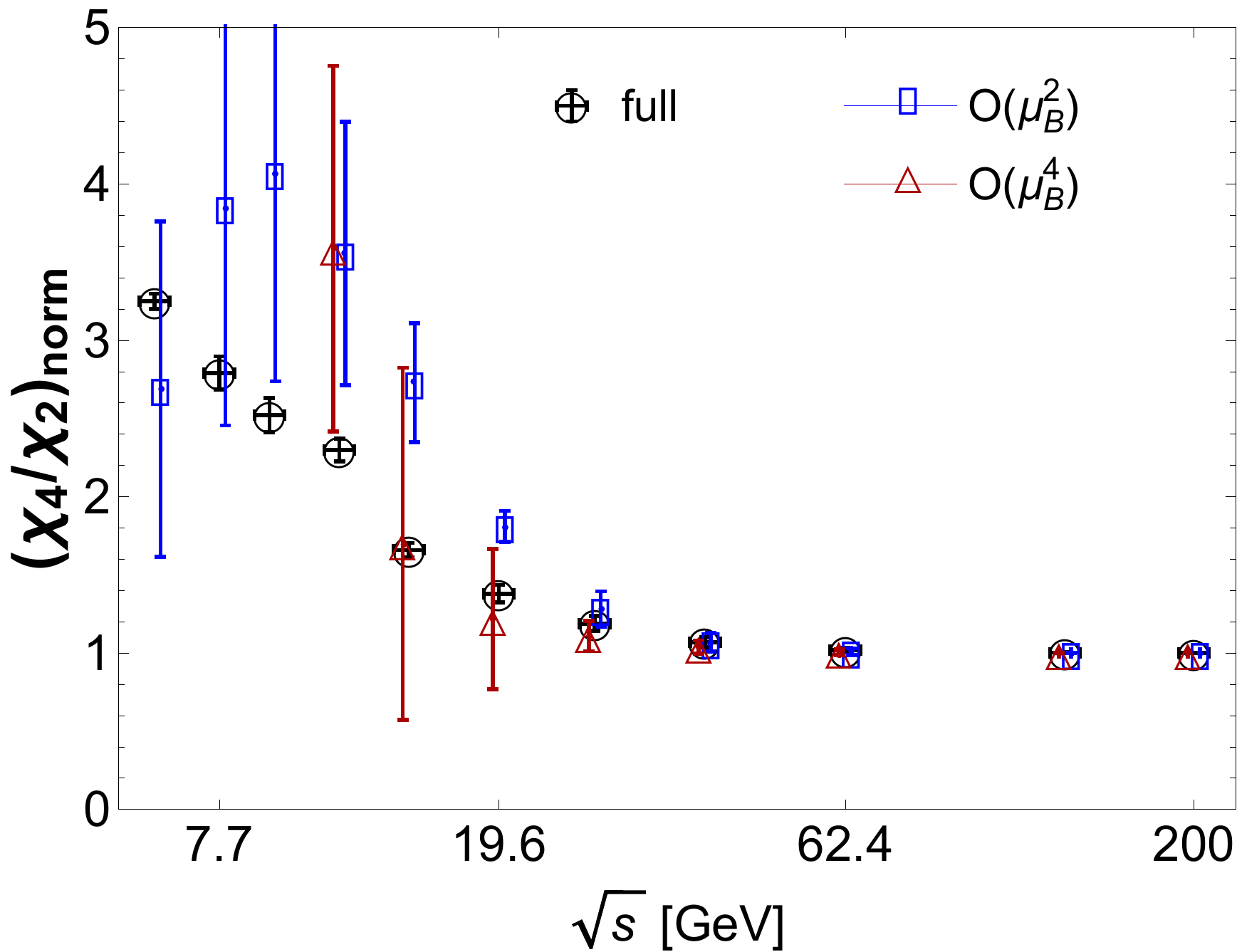} 
	\caption{\label{chi42_recon}(Color online) Ratio between $\chi_4$ and $\chi_2$ (normalized by its value at $\sqrt{s}=200$ GeV) along the transition line trajectory (shown in Fig.\ \ref{snn} of the main text). A comparison is shown between the full result for this ratio computed using black hole engineering and the corresponding reconstructed results using a Taylor series including terms up to $\mathcal{O}(\mu_B^2)$ and $\mathcal{O}(\mu_B^4)$.}
\end{figure}

Interestingly enough, the reconstructed ratio $\chi_4/\chi_2$ (normalized by its value at $\sqrt{s}=200$ GeV), along the transition lines defined near Fig.\ \ref{snn} of the Methods, works very well using terms up to $\mathcal{O}(\mu_B^4)$ down to energies as low as $\sqrt{s}=14.5$ GeV, as shown in Fig.\ \ref{chi42_recon}.  However, it is also clear from Fig.\ \ref{chi42_recon} that a truncation at $\mathcal{O}(\mu_B^2)$ can only reasonably reconstruct the ratio $\chi_4/\chi_2$ in an extremely limited range, which is not enough to cover all the values of $\sqrt{s}$ probed in the Beam Energy Scan at RHIC. While the inclusion of terms up to $\mathcal{O}(\mu_B^4)$ can be used to determine the $\chi_4/\chi_2$ ratio at energies as low as $\sqrt{s}=14.5$ GeV, Fig.\ \ref{chi42_recon} clearly demonstrates that lattice QCD calculations will need to determine this ratio at least up to $\mathcal{O}(\mu_B^6)$ or higher to describe large values of $\chi_4/\chi_2$ at low $\sqrt{s}$.

\section{Analysis of the radius of convergence of the Taylor series}
\label{sec:rad}

Since we know where the critical point of the black hole engineering model is located in the phase diagram, one can use this information to determine that the radius of convergence of the Taylor series is $\mu_B^{CEP}/T_{CEP}\sim 8.1$, assuming that the critical point is the first singularity encountered increasing $\mu_B/T$. This allows us to directly check some methods to determine the radius of convergence using the results from the low order baryon susceptibilities. For instance, in the recent paper \cite{DElia:2016jqh}, estimates for the radius of convergence were made using the quantities, 
\begin{equation}
\rho^f_{n,m}(T)=\left(\frac{\frac{m!}{n!}\chi_n(T)}{\chi_m(T)}\right)^{1/(m-n)}
\end{equation}
and
\begin{equation}
\rho^{\chi}_{n,m}(T)=\left(\frac{\frac{(m-2)!}{(n-2)!}\chi_n(T)}{\chi_m(T)}\right)^{1/(m-n)}
\end{equation}
obtained from a power series expansion in $\mu_B/T$ of the pressure and the baryon density. These estimates necessarily coincide when $n$ and/or $m$ go to infinity, giving the radius of convergence of the series for a given $T$. However, considering that only a few terms of the series are known in practice, the expectation is that a consistent determination of the critical point appears when the estimators above agree with each or show some sign of convergence.
  
\begin{figure*}
	\begin{center}
		\includegraphics[width=35pc]{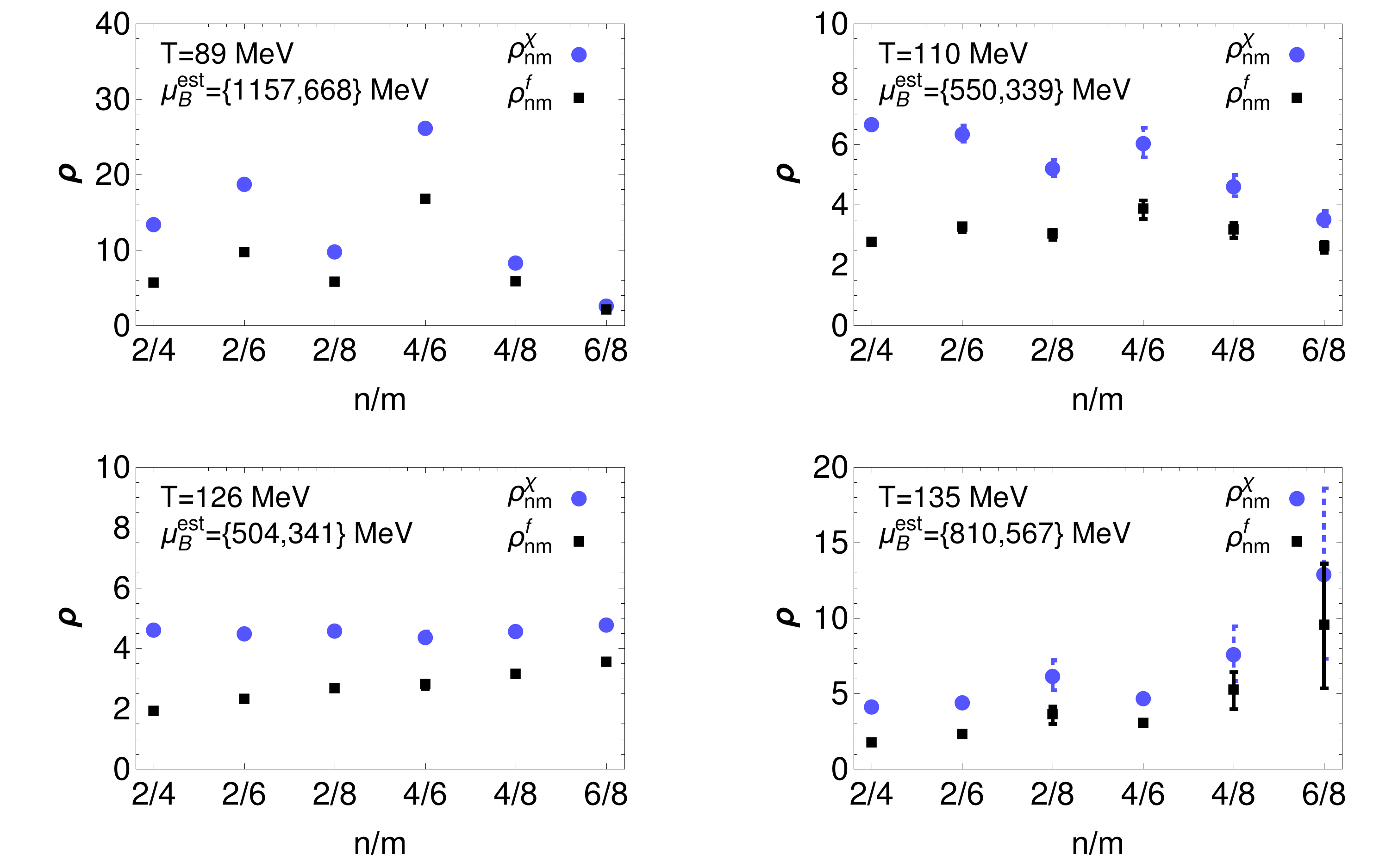}
	\end{center}
    \caption{\label{Fig:rhogrid}(Color online) Estimates for the radius of convergence, $\rho^f_{n,m}$ and $\rho^{\chi}_{n,m}$, defined using the Taylor series expansions for the pressure and the baryon density, as recently studied in \cite{DElia:2016jqh}, including terms up to $(n,m)=8$ for our black hole model with $T=89$, 110, 126, and 135 MeV. For the $T=89$ MeV case, which corresponds to the value of the critical temperature of the CEP of the model, only the upper bounds of $\chi_6$ and $\chi_8$ are taken into account.}
\end{figure*}

As a reminder, our critical point is at $\left(T_{CEP}=89,\mu_B^{CEP}=724\right)$ MeV so one may investigate if, at $T=89$ MeV, $\rho^f_{n,m} \sim \rho^{\chi}_{n,m}\sim 8.1$ for the largest values of $n$ and $m$. Even though a calculation of the susceptibilities at $T=89$ MeV (and $\mu_B=0$) is numerically challenging due to the extremely small size of $\chi_2$ at low temperatures, one can at least use an upper bound for the susceptibilities (see Methods), which is shown in Fig.\ \ref{Fig:rhogrid} (top, left) together with the corresponding (extremely) rough estimate of $\rho^{\chi}_{n,m}\sim 13$, which gives $\mu_B^{est} \sim 1157$ MeV and $\rho^f_{n,m}\sim 7.5$, which then gives $\mu_B^{est} \sim 668$ MeV.  While $\rho^f_{n,m}$ is closer to the true value for the radius of convergence of our critical point, we note that the lower bound of the higher order susceptibilities could not be included here due to numerical difficulties. More importantly, one can clearly see that there is a large variation in $\rho^f_{n,m}$ and $\rho^{\chi}_{n,m}$ even for large values of $(n,m)$, which shows that no convergence has been observed yet.

The other difficulty that is inherent in these calculations is that the critical temperature $T_{CEP}$ is still unknown on the lattice. Thus, if one were to scan other temperatures, one could actually receive false positives where $\rho^f_{n,m} \sim \rho^{\chi}_{n,m} \sim constant$ for the largest values of $n$ and $m$ available.  For instance, in Fig.\ \ref{Fig:rhogrid} we show $\rho^{\chi}_{n,m}$ and $\rho^f_{n,m}$ for all combinations of $(n,m)$ up to 8 at $T=110$, 126, and 135 MeV. The choice $T=135$ MeV was motivated by the results shown in \cite{DElia:2016jqh} at the same temperature and, in fact, we find that in our model the ratios have both the same order of magnitude and the same qualitative behavior found in \cite{DElia:2016jqh}.  At $T=110$ MeV our model exhibits a somewhat flat behavior that leads to the estimate $ \rho^f_{n,m}\sim 3$ such that $\mu_B^{est}\sim 339$ MeV. The case where $T=126$ MeV is the most interesting since both estimators are nearly flat and they seem to begin to converge with $\rho^f_{n,m}\sim 4$ and $\rho^{\chi}_{n,m}\sim 2.7$, which give $\mu_B^{est}=504$ MeV and $\mu_B^{est}=341$ MeV, respectively. In the absence of previous knowledge of the exact critical temperature, the four temperature radius of convergence scan using $(n,m)$ up to 8 shown in \ref{Fig:rhogrid} would lead to the wrong conclusion that $T_{CEP}$ is closer to $T=126$ MeV than to its actual value of 89 MeV.

\begin{figure}
	\begin{center}
		\includegraphics[width=20pc]{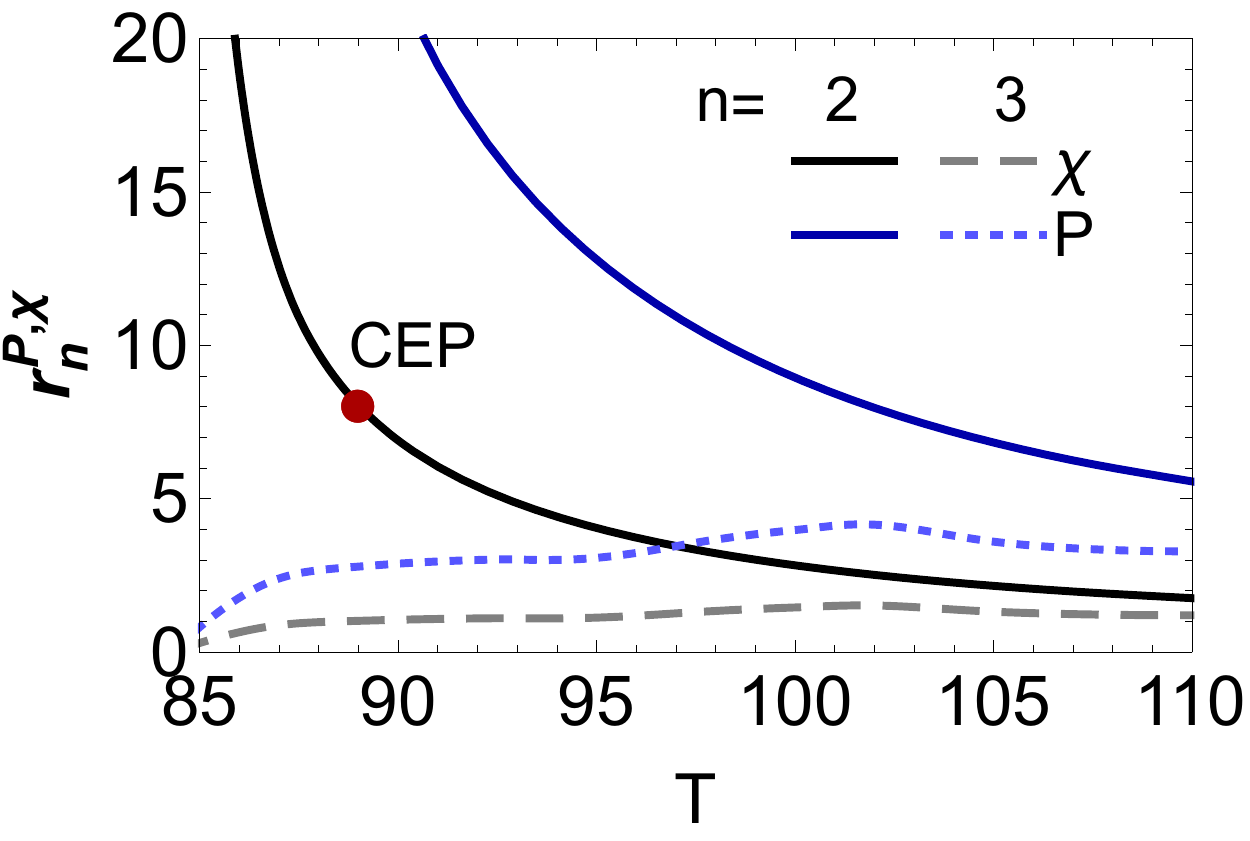}
	\end{center}
    \caption{\label{Fig:rpres}(Color online) Estimates for the radius of convergence, $r^P_{n}$ and $r^{\chi}_{n}$, used in \cite{Bazavov:2017dus}, considering terms up to $n=3$ using our black hole model (only the upper bounds of $\chi_6$ and $\chi_8$ are shown).}
\end{figure}

Another expression was recently employed in \cite{Bazavov:2017dus} to study the radius of convergence using ratios of baryon number susceptibilities defined as
\begin{equation}
r^P_{n}(T)=\left|\frac{(2n+2)(2n+1)\chi_{2n}(T)}{ \chi_{2n+2}(T)}\right|^{1/2} 
\end{equation}
for the pressure series, while for the baryon density series one writes 
\begin{equation}
r^{\chi}_{n}(T)=\left|\frac{2n(2n+1)\chi_{2n}(T)}{ \chi_{2n+2}(T)}\right|^{1/2}.
\end{equation}
Again, the radius of convergence of both series are the same being formally determined by the limit $n\rightarrow \infty$. The results of such an analysis involving the higher order baryon susceptibilities of our model are shown in Fig.\ \ref{Fig:rpres}.  Again, due to the critical point being at a low value in temperature we are only able to include the upper bound of our $\chi_6$ and  $\chi_8$. Up to $n=3$ one finds $r^P_{n}\sim 2.7 r^{\chi}_{n}$, however, the results are certainly closer to each other than for $n=2$.  Additionally, $r^{\chi}_{n}$ computed at $n=2$ is found to be very close to the correct radius of convergence. However, this is most likely just a coincidence since the inclusion of terms up to $n=3$ changes the result dramatically.

\begin{figure}
	\begin{center}
		\includegraphics[width=20pc]{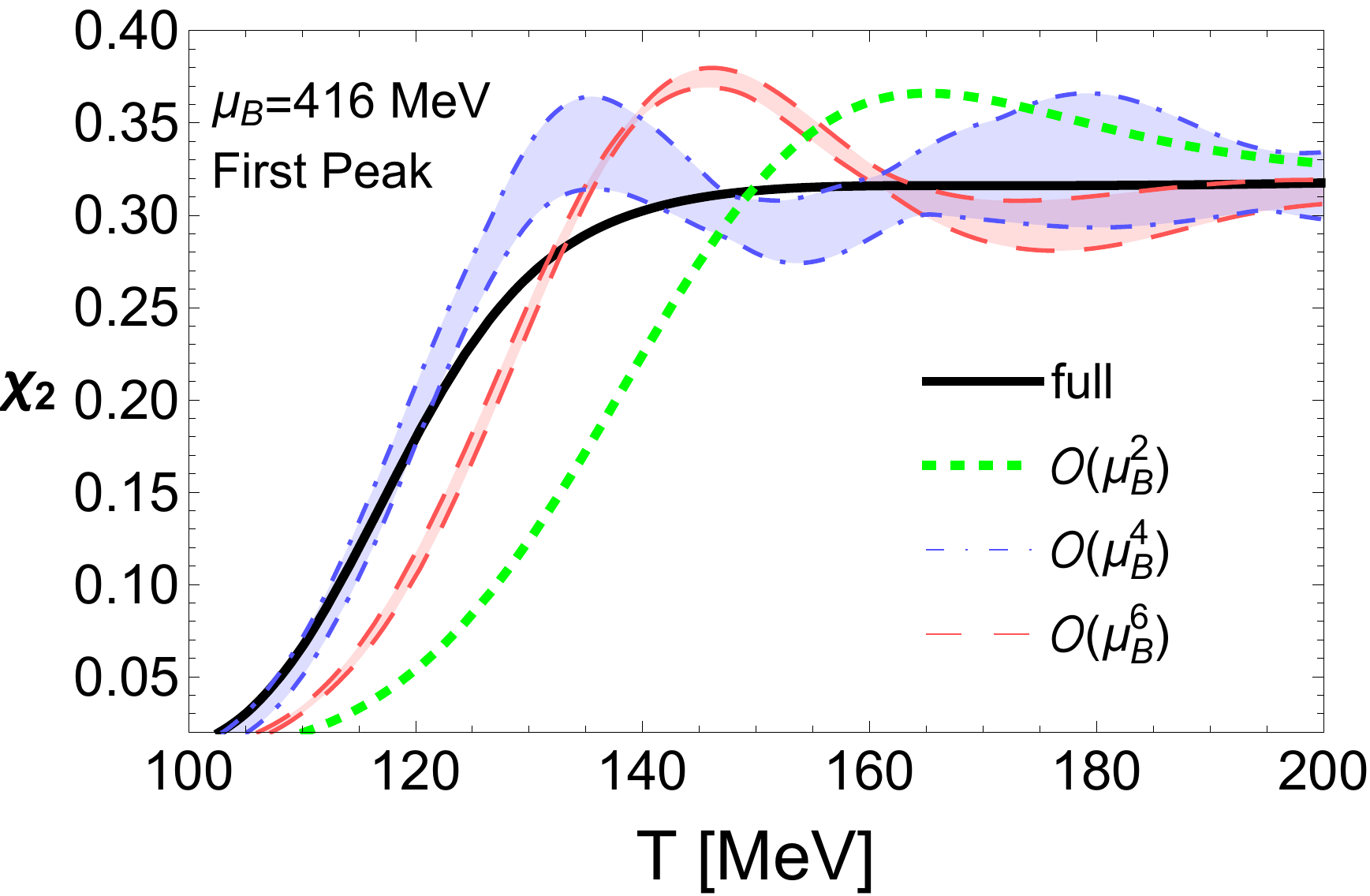}
	\end{center}
    \caption{\label{Fig:firstpeak}(Color online) $\chi_2(T,\mu_B=416 \,\mathrm{MeV})$ at the smallest value of $\mu_B$ where a maximum in $\chi_2$ is numerically found compared to the Taylor series reconstructed  $\chi_2(T,\mu_B=416\,\mathrm{MeV})$ including terms of order $\mathcal{O}(\mu_B^{2})$, $\mathcal{O}(\mu_B^{4})$, and  $\mathcal{O}(\mu_B^{6})$.}
\end{figure}

\begin{figure*}
	\begin{center}
		\includegraphics[width=35pc]{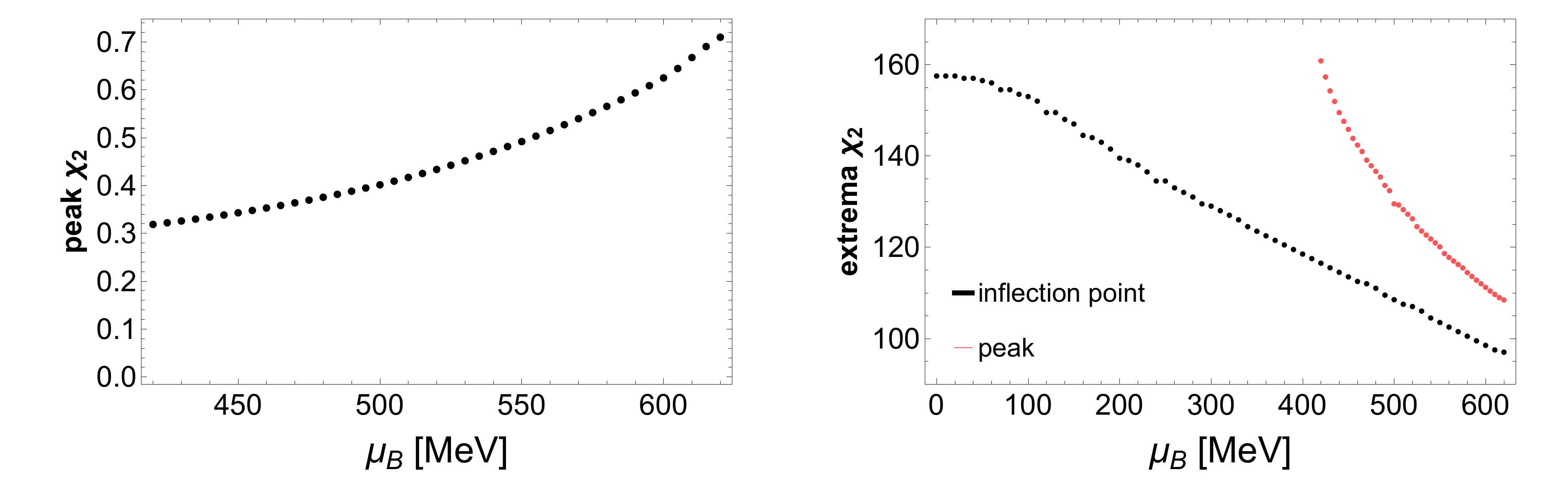}
	\end{center}
    \caption{\label{Fig:peakgrowth}(Color online) Left: Growth of the peak in $\chi_2$ across $\mu_B$. Right: Comparison of the temperature of the inflection point of $\chi_2$ across $\mu_B$ (black curve) vs. the temperature defined by the peak of $\chi_2$ across $\mu_B$ (red curve).}
\end{figure*}

Our results indicate that, unfortunately, higher order terms would be needed for this estimate of the radius of convergence to be applicable (a similar conclusion was found in other models \cite{Karsch:2010hm}).  However, we do find the close convergence of  $r^P_{n}$ to $r^{\chi}_{n}$ for $n=3$ to be promising. One should note that the previously mentioned lattice QCD results for the radius of convergence focused on values of $T$ significantly larger than our $T_{CEP}=89$ MeV. Due to the numerical difficulties faced by lattice QCD calculations at low temperatures, if the QCD critical point is indeed located at similar values of $T$ as our $T_{CEP}$, it would be a challenge for lattice QCD practitioners to employ this approach at low enough temperatures using high enough orders of baryon susceptibilities to realistically locate the critical point.

Perhaps a better indicator of the presence of a critical point is the formation of a peak in $\chi_2$, which would eventually evolve into a divergence at large enough $\mu_B$.  By taking different slices of $\chi_2(T,\mu_B=constant)$, one can find the location of the maximum of $\chi_2$ with respect to $T$. In our calculations a maximum in $\chi_2$ occurs first at $\mu_B=416$ MeV and $T=164.5$ MeV, which is shown in Fig.\ \ref{Fig:firstpeak}. The advantage of looking for the development of a maximum is that it can occur at higher temperatures that are more easily calculated within Lattice QCD. Unfortunately, even with the inclusion of terms up to $\mathcal{O}(\mu_B^{6})$ it is not yet possible to reasonably reconstruct $\chi_2(T,\mu_B=416\,\mathrm{MeV})$. However, this may be possible already with the addition of higher order terms such as $\mathcal{O}(\mu_B^{8})$ or $\mathcal{O}(\mu_B^{10})$. Of course, one could question if a peak could be formed in $\chi_2(T,\mu_B=constant)$ that either remains constant across $\mu_B$ or eventually disappears (not leading to a divergence). This possibility does not occur in our calculations and, to the best of our knowledge, this is not displayed in other effective models of QCD at finite temperature and density.

Due to the divergence of $\chi_2$ at the critical point, the peak of $\chi_2$ and its inflection point must eventually converge. In Fig.\ \ref{Fig:peakgrowth} (left) we first determine the growth of $\chi_2$ at its peak, which is shown to increase more and more quickly as $\mu_B$ increases.  We then compare the difference between the position of the inflection point and the peak of  $\chi_2$ in Fig.\ \ref{Fig:peakgrowth} (right). Indeed, we see that they converge quickly as one approaches the critical point.  Numerically, these quantities are difficult to calculate precisely close to the critical point but up to $\mu_B=625$ MeV we already see a clear convergence, expected to continue towards larger values of $\mu_B$.

\bibliographystyle{apsrev4-1}
\bibliography{BH_Bib} 

\end{document}